\documentclass[a4paper,fleqn,usenatbib]{mnras}
\usepackage{newtxtext,newtxmath}
\usepackage[T1]{fontenc}
\usepackage{ae,aecompl}
\usepackage{graphicx}   
\usepackage{amsmath}    
\usepackage{amssymb}    

\title[Susceptibility of planetary atmospheres to mass loss and growth by planetesimal impacts]
{Susceptibility of planetary atmospheres to mass loss and growth by planetesimal impacts: the impact shoreline}
\author[M. C. Wyatt et al.]
  {M. C. Wyatt$^1$\thanks{Email: wyatt@ast.cam.ac.uk},
  Q. Kral$^2$,
  C. A. Sinclair$^1$ \\
  $^1$ Institute of Astronomy, University of Cambridge, Madingley Road,
  Cambridge CB3 0HA, UK \\
  $^2$ LESIA, Observatoire de Paris, Universit\'{e} PSL, CNRS, Sorbonne Universit\'{e}, Univ. Paris Diderot, Sorbonne Paris Cit\'{e},
  5 place Jules Janssen, 92195 Meudon, France
}

\pubyear{2019}

\begin{document}
\label{firstpage}
\pagerange{\pageref{firstpage}--\pageref{lastpage}}
\maketitle

\begin{abstract}
This paper considers how planetesimal impacts affect planetary atmospheres.
Atmosphere evolution depends on the ratio of gain from volatiles to loss from atmosphere stripping
$f_{\rm v}$;
for constant bombardment, atmospheres with $f_{\rm v}<1$ are destroyed in finite time, but
grow linearly with time for $f_{\rm v}>1$.
An impact outcome prescription is used to characterise how $f_{\rm v}$ depends on
planetesimal impact velocities, size distribution and composition.
Planets that are low mass and/or close to the star have atmospheres that deplete in impacts,
while high mass and/or distant planets grow secondary atmospheres.
Dividing these outcomes is an $f_{\rm v}=1$ impact shoreline analogous to Zahnle \& Catling's cosmic shoreline.
The impact shoreline's location depends on assumed impacting planetesimal properties,
so conclusions for the atmospheric evolution of a planet like Earth with $f_{\rm v} \approx 1$
are only as strong as those assumptions.
Application to the exoplanet population shows the gap in the planet radius distribution
at $\sim 1.5R_\oplus$ is coincident with the impact shoreline, which has a similar
dependence on orbital period and stellar mass to the observed gap.
Given sufficient bombardment, planets below the gap would be expected to lose their
atmospheres, while those above could have atmospheres enhanced in volatiles.
The level of atmosphere alteration depends on the total bombardment a planet experiences, and so
on the system's (usually unknown) other planets and planetesimals,
though massive distant planets would have low accretion efficiency.
Habitable zone planets around lower luminosity stars are more susceptible to atmosphere stripping,
disfavouring M stars as hosts of life-bearing planets if Earth-like bombardment is conducive
to the development of life.
\end{abstract}


\section{Introduction}
\label{s:intro}
There are now over four thousand known exoplanets.
Many are seen to transit in front of their host stars enabling study of their atmospheres.
Atmosphere characterisation is possible not only for gas giant planets \citep{Charbonneau2002},
but also for Earth-sized planets in the habitable zone \citep[e.g.,][]{deWit2018}.
Characterisation of exoplanet atmospheres is expected to become easier as
planets are found to transit around brighter stars \citep[e.g.,][]{Rauer2014, Huang2018},
and it is within our reach to search for evidence of extraterrestrial life in exoplanet atmosphere
observations \citep[e.g.,][]{Kaltenegger2017, Defrere2018}.
As such it is important to understand the processes responsible for the origin and evolution
of planetary atmospheres \citep{Kasting2003}.
Not only will this help with the interpretation of exoplanet atmosphere observations,
in turn constraining those formation and evolution processes, but also allow consideration
of issues such as how conditions on planetary surfaces evolve.
It is not yet fully understood how these processes played out on the Solar system's terrestrial
planets \citep[e.g.,][]{Prinn1987, Lammer2018}, which nevertheless provide valuable constraints,
particularly in the regime of low mass and/or habitable planets.

In general it might be considered that a planet could acquire an atmosphere during its formation,
by accretion of either gas from the protoplanetary disk \citep[predominantly H or He,][]{Lammer2014}, or of solids
containing volatiles (such as water or CO$_2$) that are outgassed during accretion \citep{ElkinsTanton2008}.
That atmosphere could then evolve due to internal processes, such as the dissipation of the planet's initial
gravitational and thermal energy (which promote atmospheric mass loss), outgassing of volatiles
originally locked within the planet \citep{Craddock2009, ElkinsTanton2012, Godolt2019}, and geological processes such as the subduction
of CO$_2$ \citep[e.g.,][]{Walker1981, Zahnle2007}.
External processes could also be at play, such as
irradiation by the central star (that also promotes atmospheric mass loss through photoevaporation) and
impacts from planetesimals (which can both strip the atmosphere and deliver volatiles to it).

The broad properties of the exoplanet population can be explained with a subset of the processes mentioned above. 
For example, the core accretion paradigm in which giant planets accrete significant atmospheres once their
cores reach $\gg 1-10M_\oplus$ \citep{Pollack1996, Brouwers2018} is successful at explaining the distribution of planetary masses
and radii \citep[e.g.,][]{Jin2018}.
Planets smaller than $1.6R_\oplus$ are inferred to have (at most) tenuous atmospheres, while those up to $\sim 4R_\oplus$ have
atmospheres with a few \% by mass, though there is some degeneracy when inferring atmosphere mass depending on
whether the planet's mass is dominated by volatiles \citep{Rogers2015, Lozovsky2018}.
There is direct evidence for photoevaporative mass loss in some systems \citep[e.g.,][]{VidalMadjar2003}, which when applied
to the broader population can explain the absence of large planets at small orbital distances \citep[e.g.,][]{Lecavelier2007}.
Most recently a gap in the distribution of planetary radii at $\sim 1.5R_\oplus$ \citep{Fulton2017,VanEylen2018} has also been explained by
photoevaporation by stellar X-rays that are prevalent during the first 100\,Myr or so of a star's life \citep{Jackson2012};
more massive atmospheres are not lost on this timescale and so can be retained, while those below this
level are destroyed \citep{Owen2017, Lehmer2017}.
An alternative explanation for this gap has also been given as mass loss driven by the luminosity
of the cooling core \citep{Ginzburg2018}.

A similar story applies to the planets and moons in the Solar system, for which the
presence or absence of an atmosphere is determined by the ratio of insolation to escape
velocity to the fourth power, creating a {\it cosmic shoreline} that may be explained by
hydrodynamic thermal escape or irradiation \citep{Zahnle2017}.
However, Solar system studies also highlight the potential contribution of impacting planetesimals \citep[e.g.,][]{Cameron1983,
Ahrens1993}.
For example, \citet{Zahnle1992} noted that the difference between Titan's atmosphere and the lack of one on Ganymede
and Callisto could be explained by the lower impact velocity onto Titan which can thus retain an impact
generated atmosphere, and impacts are thought to be responsible for the erosion of Mars' primordial atmosphere 
\citep{Melosh1989}.
Indeed \citet{Zahnle2017} note that the cosmic shoreline may alternatively be explained by impact erosion,
but do not consider that possibility in as much detail because of uncertainties in how to model this.
The Earth's atmosphere is also thought to have been affected by impacts, having its origin
in a combination of gas from the protosolar nebula and accreted cometary volatiles \citep{Owen1995, Dauphas2003},
with impacts also postulated as the origin of
the Earth's oceans \citep{Chyba1990a}, as well as a means of delivering organic molecules \citep{Chyba1990b}.
This interpretation is however challenged by the detailed volatile compositions of Earth and comets,
which suggest that comets are not the dominant reservoir \citep{Marty2016},
though the picture for noble gases is more complicated \citep{Marty2017, Zahnle2019}.

Clearly there are many competing processes that affect atmosphere evolution.
This paper focusses on one of those processes, which is the effect of planetesimal impacts, both
their role in stripping a pre-existing atmosphere, and in delivering volatiles to replenish that atmosphere.
These processes have previously been applied to consideration of the evolution of Solar system terrestrial
planets \citep[e.g.,][]{Melosh1989, Svetsov2007, deNiem2012, Schlichting2015, Pham2016}.
However, there are differences in the prescriptions for the outcome of collisions between these studies,
as well as in their assumptions about the impactors, which lead to slightly different conclusions.
These studies are nevertheless converging on the most appropriate prescription,
with analytical considerations of the underlying physics of impacts \citep{Schlichting2015}
in broad agreement with numerical simulations \citep{Shuvalov2009}, 
for example in the conclusion that mass loss should be dominated by impacts with
planetesimals a few km in size.
Giant impacts are generally considered to play a less significant role in atmosphere evolution
\citep[e.g.,][]{Genda2003,Schlichting2018}, though these can provide an element of stochasticity
to explain different atmosphere properties seen in the same system \citep{Griffith1995, Biersteker2018}, 
could be more important for planets with oceans \citep{Genda2005},
and may promote degassing explaining some features of the atmosphere of Venus \citep{Gillmann2016}.

While the parameterisation of \citet{Shuvalov2009} can be extended across a wide range of parameter space,
these models for impact driven atmosphere evolution have not yet been applied to the broader
range of planets in the exoplanet population, except in the case of the TRAPPIST-1 planetary system \citep{Kral2018}.
This paper aims to address exactly this topic, for example to consider the possibility of
an impact shoreline that determines whether planets (and moons) have an atmosphere.
It starts in \S \ref{s:simp} by considering how atmospheres evolve with a very basic 
prescription for the outcome of impacts.
The simulations of \citet{Shuvalov2009} are then used in \S \ref{s:comp} to develop a more detailed
model which is applied to atmospheres across a broad range of planet mass and distance from stars of
different types.
The results are summarised in \S \ref{s:disc}, where the model is also applied to the exoplanet population
to consider what effect impacts may have had on their observable properties, and to the Solar system
planets to consider how conclusions for atmosphere evolution depend on assumptions about the impacting planetesimals.

\section{Simple atmosphere evolution model}
\label{s:simp}
Consider a model in which a planet's atmosphere has a total mass $m=m_{\rm p}+m_{\rm v}$ which
is made up of a primordial component ($m_{\rm p}$) and a volatile component ($m_{\rm v}$) that is delivered
later (to replenish a secondary atmosphere) by planetesimal impacts that also lead to atmospheric mass loss.
We will assume that atmospheric mass is lost at a rate $\dot{m}^{-}$ and that volatiles are delivered
at a rate $\dot{m}^{+}_{\rm v}$, so that $\dot{m} = \dot{m}^{+}_{\rm v} - \dot{m}^{-}$ and $\dot{m}_{\rm p} = -(m_{\rm p}/m)\dot{m}^{-}$.

If both of these rates are constant the resulting evolution of the atmospheric mass is
\begin{eqnarray}
  m/m_0         & = &  1 + (f_{\rm v}-1)t/t_0, \label{eq:mm0simp} \\
  m_{\rm p}/m_0 & = &  (m/m_0)^\frac{1}{1-f_{\rm v}}, \label{eq:mpm0} 
\end{eqnarray}
and $m_{\rm v}=m-m_{\rm p}$, where $m_0$ is the initial atmospheric mass (all of which is primordial), $f_{\rm v}=\dot{m}^{+}_{\rm v}/\dot{m}^{-}$ is the
ratio of atmospheric mass gain and loss rates, and $t_0=m_0/\dot{m}^{-}$ is the time it would take to deplete the primordial atmosphere
in the absence of any gain from volatile delivery.

\begin{figure}
  \begin{center}
    \vspace{-0.1in}
    \begin{tabular}{c}
      \hspace{-0.55in} \includegraphics[width=1.25\columnwidth]{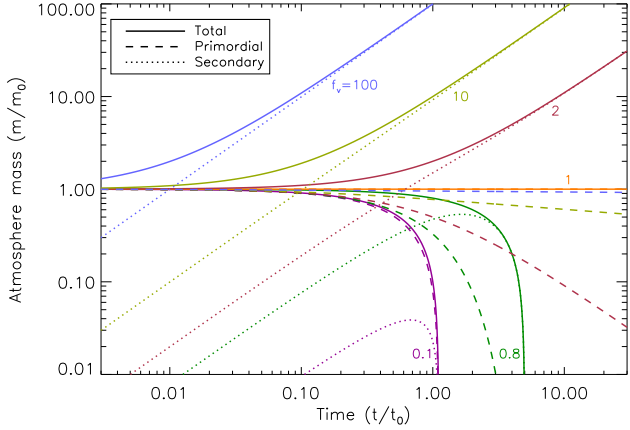}
    \end{tabular}
    \vspace{-2.55in}
    \caption{Simple model for the evolution of atmospheric mass in which the rates of gain (due to accretion of volatiles that replenishes a
    secondary atmosphere) and loss (that depletes both primordial and secondary atmospheres) are assumed to be constant.
    The evolution depends only on the ratio of the gain and loss rates given by the parameter $f_{\rm v}$, six different values of which
    are shown with different colours as noted in the annotation.
    The solid lines show the total atmospheric mass, which is made up of a primordial component shown with dashed lines and a secondary component
    shown with dotted lines.
    }
   \label{fig:simp}
  \end{center}
\end{figure}

The evolution from this simple model is plotted in Fig.~\ref{fig:simp} for a range of its only free parameter $f_{\rm v}$.
While this oversimplifies the problem, since these rates ($\dot{m}^{+}_{\rm v}$ and $\dot{m}^{-}$)
are expected to have a dependence on atmospheric mass which is itself
varying, it serves to illustrate an important point.
This is that the evolution depends critically on the parameter $f_{\rm v}$ which determines whether, overall, atmospheres gain
or lose mass in planetesimal collisions.
If they gain mass (i.e., if $f_{\rm v}>1$) then atmospheres grow linearly with time for $t/t_0 \gg 1$
becoming dominated by the secondary component (see e.g. the $f_{\rm v}=2$, 10 and 100 lines on Fig.~\ref{fig:simp}).
If on the other hand they lose mass (i.e., if $f_{\rm v}<1$) then while the secondary component starts to grow
in mass, this growth will eventually be reversed and the whole atmosphere will deplete to
zero in a finite time (see e.g. the $f_{\rm v}=0.1$ and 0.8 lines on Fig.~\ref{fig:simp}).
Either way the atmospheric composition becomes more volatile rich with time.

\section{Physically based atmosphere evolution model}
\label{s:comp}
The model of \S \ref{s:simp} can be improved using a prescription for the outcome of planetesimal impacts.
Here, similar to \citet{Kral2018}, we use the results of \citet{Shuvalov2009}
which considered simulations of planetesimals of sizes in the range $1-30$\,km
impacting at $10-70$\,km\,s$^{-1}$ onto planets that have Earth-like atmospheres.
These results can be scaled to arbitrarily large or small impactors, impact velocities and atmospheric densities,
in a way that can be understood within a framework that describes the underlying physics \citep{Schlichting2015}.
However, care is needed when applying the results outside the range of the original simulations, since the relevant
physics may be different for impacts in different regimes.
In particular, the \citet{Shuvalov2009} prescription is only valid for impactors that reach a planet's surface causing a
cratering-like event and local atmospheric mass loss, whereas for small impactors, or for those interacting with very
dense atmospheres, the impactors can be decelerated and may fragment or undergo an aerial burst before reaching the surface.
At the other extreme, massive impactors can send shock waves through the planet causing non-local
atmospheric loss, which is not accounted for by \citet{Shuvalov2009}.
Aerial bursts have been studied \citep[e.g.,][]{Shuvalov2014}, but the prescriptions that are available are not general
enough to be useful for the current study, and so such effects are ignored for now, and this caveat will be
discussed further in \S \ref{ss:massive}.
Giant impacts are discussed in \S \ref{ss:stochasticity} where it is shown that they only become important for atmospheres
that contain a substantial fraction of the planet's mass.

\subsection{Assumptions about planet atmosphere}
\label{ss:atm}
The starting point of the model is to define the planet's atmosphere, which is assumed to be isothermal at temperature
$T=278L_\star^{1/4}a_{\rm p}^{-1/2}$\,K,
where $L_\star$ is the stellar luminosity in units of ${\rm L}_\odot$ and $a_{\rm p}$ is the semimajor axis in au of the planet's orbit
(which is assumed to be circular).
The parameters used in this paper and their units are summarised in Table~\ref{tab:units}.
This temperature sets the scale height of the planet's atmosphere $H=kT/(\mu m_{\rm H}g)$, where $k$ is Boltzmann's constant, $\mu$ is the mean
molecular weight of the atmosphere, $m_{\rm H}$ is the mass of Hydrogen, and $g=GM_{\rm p}R_{\rm p}^{-2}$ is the planet's surface gravity,
$M_{\rm p}$ is the planet's mass (which will be in ${\rm M}_\oplus$ throughout) and $R_{\rm p}$ its radius (at the solid surface).
Note that later equations will be expressed in terms of the planet's mass and mean density ($\rho_{\rm p}$),
rather than its mass and radius (these quantities being related by assuming a spherical planet).
Later plots will also consider planet density to be $\rho_{\rm p}=\rho_\oplus=5.5$\,g\,cm$^{-3}$,
though we might equally have included a dependence on mass or composition
\citep[e.g., from][a dependence of $\rho_{\rm p} \propto M_{\rm p}^{0.19-0.25}$ can be inferred]{Lopez2014, Zeng2016}.
We will consider two bounding cases for $\mu$, which is that of a primordial (solar) composition $\mu_{\odot}=2.35$, and that
of a volatile-rich (Earth-like) composition $\mu_\oplus=29$.

Combining these assumptions gives for the atmospheric scale height
\begin{equation}
  H=H_0 L_\star^{1/4} a_{\rm p}^{-1/2} M_{\rm p}^{-1/3} \rho_{\rm p}^{-2/3} \mu^{-1}, \label{eq:h}
\end{equation}
where $\rho_{\rm p}$ is the planet's density in g\,cm$^{-3}$,
and $H_0=0.73 \times 10^6$\,m (meaning that these assumptions give $H_\oplus=8100$\,m for the Earth).
We will assume $H \ll R_{\rm p}$ throughout, which for the given assumptions means that the results are applicable
to planets with $M_{\rm p} \gg 0.017 L_\star^{3/8} \rho_{\rm p}^{-1/2} \mu^{-3/2} a_{\rm p}^{-3/4}$;
this only excludes extremely low mass planets that are very close to the star, which are not seen yet in the exoplanet
population and are not considered here.
This means that the total atmospheric mass ($m$) scales with the atmospheric density at the planet's surface ($\rho_0$)
according to $m \approx 4\pi H R_{\rm p}^2\rho_0$, where for the Earth $m_\oplus=0.85 \times 10^{-6}{\rm M}_\oplus$.
In some of the analysis the atmosphere mass will be defined by its ratio to the planet mass, $\delta=m/M_{\rm p}$, with
atmospheres starting out with a mass $m_0 = \delta_0 M_{\rm p}$, and the Earth having $\delta_\oplus=0.85 \times 10^{-6}$.
The above assumptions also mean that the pressure at the planet's surface is
\begin{equation}
  p/p_\oplus = (\rho_{\rm p}/\rho_\oplus)^{4/3} (M_{\rm p}/{\rm M}_\oplus)^{2/3} (\delta/\delta_\oplus),
  \label{eq:p}
\end{equation}
where $p_\oplus$ is the pressure at the Earth's surface.

For atmospheres significantly more massive than that of the Earth the assumption that they are isothermal is no longer
valid.
The outermost regions will still be isothermal for such atmospheres, but there is a significant portion below this which
may be adiabatic down to the surface.
While simple prescriptions for the structure of such atmospheres exist \citep[e.g.,][]{Owen2017}, here we prefer to leave
consideration of massive atmospheres, such as those with $\delta \approx 1$\% seen in the transiting exoplanet
population \citep[e.g.,][]{Wolfgang2015,Fulton2017}, to a future study.

\subsection{Outcome of individual impacts}
\label{ss:ind}
The outcome of a collision with a planetesimal of diameter $D$ and density $\rho_{\rm imp}$ at an impact velocity $v_{\rm imp}$
is determined by the dimensionless parameter \citep[called erosional efficiency by][]{Shuvalov2009}
$\eta=(D/H)^3[(v_{\rm imp}/v_{\rm esc})^2-1][\rho_{\rm imp}\rho_{\rm ps}/(\rho_0(\rho_{\rm imp}+\rho_{\rm ps}))]$, where
$v_{\rm esc}=\sqrt{2GM_{\rm p}/R_{\rm p}}$ is the planet's escape velocity, and $\rho_{\rm ps}$ is the density of the planet at its
surface which will be assumed to be equal to $\rho_{\rm p}$ (i.e., the planet is assumed to have uniform density throughout).
Given the assumptions about the planet's atmosphere in \S \ref{ss:atm} this means that
\begin{eqnarray}
  \eta & = & \eta_0 L_\star^{-1/2} a_{\rm p} M_{\rm p}^{4/3} \rho_{\rm p}^{5/3} m^{-1} \mu^2 D^3 (1+\rho_{\rm p}/\rho_{\rm imp})^{-1} \times \nonumber \\
       &   & [(v_{\rm imp}/v_{\rm esc})^2-1],
  \label{eq:eta}
\end{eqnarray}
where $\eta_0=0.5 \times 10^{-18}$ for other parameters in the units of Table~\ref{tab:units} (i.e., with $m$ in ${\rm M}_\oplus$ and $D$ in m).
For example, $\eta=8.5 \times 10^{-9}D^3$ for impacts onto the Earth with $v_{\rm imp}/v_{\rm esc}=2$ and $\rho_{\rm p}/\rho_{\rm imp}=2$
(so that the last two parentheses cancel).
According to \citet{Shuvalov2009} the atmospheric mass lost due to this impactor per impactor mass
(where $m_{\rm imp}=(\pi/6)\rho_{\rm imp}D^3$) is given by
\begin{equation}
  m_{\rm atmloss}(D)/m_{\rm imp} = [(v_{\rm imp}/v_{\rm esc})^2-1]\chi_{\rm a},
  \label{eq:matmloss}
\end{equation}
where $\log{\chi_{\rm a}} = -6.375 + 5.239\log{\eta} - 2.121(\log{\eta})^2 + 0.397(\log{\eta})^3 - 0.037(\log{\eta})^4 + 0.0013(\log{\eta})^5$
for $\log{\eta}<6$. 
To avoid the unphysical extrapolation to large $\eta$ in the parameterisation of \citet{Shuvalov2009}, we extrapolate from a fit to
their results in the range $\log{\eta}=4-6$ to find a prescription for $\log{\eta} \geq 6$ of
$\log{\chi_{\rm a}} = 0.4746 - 0.6438\log{\eta}$ that is consistent with \citet{Schlichting2015}.
The mass gain due to this impactor per impactor mass is given by
\begin{equation}
  m_{\rm impacc}(D)/m_{\rm imp} = [1-\chi_{\rm pr}],
  \label{eq:mimpacc}
\end{equation}
where
$\chi_{\rm pr} = 0$ for $\eta<10$,
$\chi_{\rm pr}= {\rm min}[0.07(\rho_{\rm p}/\rho_{\rm imp})(v_{\rm imp}/v_{\rm esc})(\log{\eta}-1) , 1]$ for $10<\eta<1000$, and
$\chi_{\rm pr}= {\rm min}[0.14(\rho_{\rm p}/\rho_{\rm imp})(v_{\rm imp}/v_{\rm esc}) , 1]$ for $\eta>1000$
\citep[$\eta>1000$ being the {\it airless} limit noted in][for which atmosphere drag is negligible for plume expansion]{Shuvalov2009}.

\begin{figure}
  \begin{center}
    \vspace{-0.1in}
    \begin{tabular}{c}
      \hspace{-0.55in} \includegraphics[width=1.25\columnwidth]{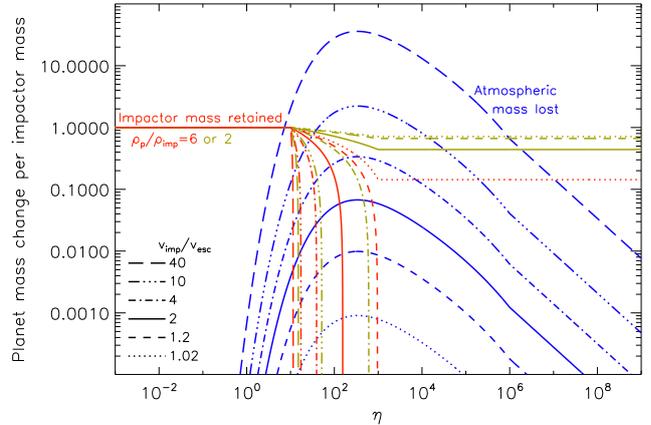}
    \end{tabular}
    \vspace{-2.55in}
    \caption{Change in planet mass due to a collision with an impactor at different levels of impact velocity relative to
    the planet's escape speed shown with different lines.
    Atmospheric mass lost per impactor mass is shown in blue, and the fraction of the impactor mass that is retained by the
    planet is shown in red or green for $\rho_{\rm p}/\rho_{\rm imp}=6$ or 2 respectively.
    The $x$-axis scales with the impactor diameter as given in eq.~\ref{eq:eta}.
    All calculations use the prescription in \citet{Shuvalov2009}.
    }
   \label{fig:mm}
  \end{center}
\end{figure}

The prescriptions from eqs.~\ref{eq:matmloss} and \ref{eq:mimpacc} are shown in Fig.~\ref{fig:mm}.
The large scale features of this figure were discussed in \citet{Shuvalov2009} and \citet{Schlichting2015}.
That is, atmospheric mass loss is most efficient for planetesimals in the middle of the size range (approximately
km-sized for Earth-like atmospheres), because larger planetesimals can only remove up to the atmospheric mass in the 
local vicinity of the impact (i.e., the polar cap), while smaller planetesimals do not impart sufficient energy to the atmosphere to remove
significant mass.
For example, Fig.~\ref{fig:mm} shows that the most erosive planetesimals for $v_{\rm imp}/v_{\rm esc}=10$ remove approximately
twice their own mass from the atmosphere.
Similarly, all of the mass of small planetesimals is retained, but for larger planetesimals much of their mass is
lost from the planet as it has too much energy to remain bound (except at very low impact velocities).
Setting eq.~\ref{eq:mimpacc} to zero shows that this transition occurs at the size for which $\eta=\eta_{\rm maxret}$, where
\begin{equation}
  \eta_{\rm maxret} = 10^{1+14(v_{\rm esc}/v_{\rm imp})(\rho_{\rm imp}/\rho_{\rm p})}
  \label{eq:etamaxret}
\end{equation}
for $v_{\rm imp}/v_{\rm esc} > 7.1\rho_{\rm imp}/\rho_{\rm p}$ (and $\eta_{\rm maxret}=\eta_{\rm max}$ for lower impact velocities).

Fig.~\ref{fig:mm} highlights that the most important free parameter that determines mass loss and gain by the planet in this
prescription is the ratio of the impact velocity to the planet's escape speed.
Larger impact velocities result in both greater levels of atmospheric mass loss and less retention of impactor mass (through a decrease
in the impactor size that can be retained).
The only other variable is the ratio of the planet's density to that of the impactor, $\rho_{\rm p}/\rho_{\rm imp}$,
which affects the impactor mass that can be retained.
Impactors that have larger densities (e.g., $\rho_{\rm p}/\rho_{\rm imp}=2$ might correspond to asteroid-like objects impacting
the Earth) can be retained up to larger sizes than those of lower densities (e.g., $\rho_{\rm p}/\rho_{\rm imp}=6$ might correspond to
comet-like objects impacting the Earth).

\subsection{Outcome of multiple impacts}
\label{ss:mult}
To determine the effect of multiple impacts onto a planetary atmosphere requires an assumption about the size distribution
of impactors.
Here we assume that there is a power law size distribution of impactors characterised by the exponent $\alpha$, such that
the number in the size $D$ to $D+dD$ is $n(D)dD$ where $n(D) \propto D^{-\alpha}$.
An infinite collisional cascade of planetesimals with dispersal threshold independent of size would be expected to
have $\alpha=3.5$ \citep{Dohnanyi1969}, but deviations from this can be expected due to size dependent strength among others things
\citep[see e.g.,][]{Wyatt2011}, so we leave this as a free parameter.
The distribution is assumed to extend from small objects of size $D_{\rm min}$ up to a size of $D_{\rm max}$.
For now we will work on the assumption that this range is large enough to have no effect on the mass budget, because mass
loss and gain is dominated by intermediate-sized planetesimals.
However, this is discussed further below, since for extreme slopes in the size distribution, or for atmospheres that
are (or become) significantly different to that of the Earth, it can be objects at the edges of the size distribution that
dominate the atmosphere's mass evolution.

While \citet{deNiem2012} found that the stochastic effect of impacts with large bodies can dominate atmospheric evolution,
we assume here that this stochasticity can be ignored, and consider that the mean change in a planet's mass
can be obtained by integrating eqs.~\ref{eq:matmloss} and \ref{eq:mimpacc} over the aforementioned size distribution
\citep[as in][]{Kral2018}. 
The possibility of stochasticity, and the effect of giant impacts more generally, is considered in \S \ref{ss:stochasticity}. 

If the total mass of impactors that collide with a planet is $m_{\rm ac}$, the atmospheric mass loss and impactor mass retained
per $m_{\rm ac}$ are
\begin{eqnarray}
  \frac{m_{\rm atmloss}}{m_{\rm ac}} & = & A \left[ \left(\frac{v_{\rm imp}}{v_{\rm esc}}\right)^2-1\right]
                                         \int_{\eta_{\rm min}}^{\eta_{\rm max}} \eta^{(1-\alpha)/3} \chi_{\rm a} d\eta, \label{eq:matmlosstot}\\
  \frac{m_{\rm impacc}}{m_{\rm ac}}  & = & A \int_{\eta_{\rm min}}^{\eta_{\rm max}} \eta^{(1-\alpha)/3} [1-\chi_{\rm pr}] d\eta, \label{eq:mimpacctot} \\
  A & = & \left( \frac{4-\alpha}{D_{\rm max}^{4-\alpha}-D_{\rm min}^{4-\alpha}} \right) \frac{1}{3} \left( \frac{\eta}{D^3} \right)^{(\alpha-4)/3}, \label{eq:a} 
\end{eqnarray}
where $\eta_{\rm min}$ and $\eta_{\rm max}$ map onto $D_{\rm min}$ and $D_{\rm max}$ through eq.~\ref{eq:eta} which is also used
to get the ratio $\eta/D^3$ in eq.~\ref{eq:a};
for the specific case of $\alpha=4$, eq.~\ref{eq:a} needs to be revised to $A=[3\ln{(D_{\rm max}/D_{\rm min})}]^{-1}$.

\begin{figure}
  \begin{center}
    \vspace{-0.1in}
    \begin{tabular}{c}
      \hspace{-0.55in} \includegraphics[width=1.25\columnwidth]{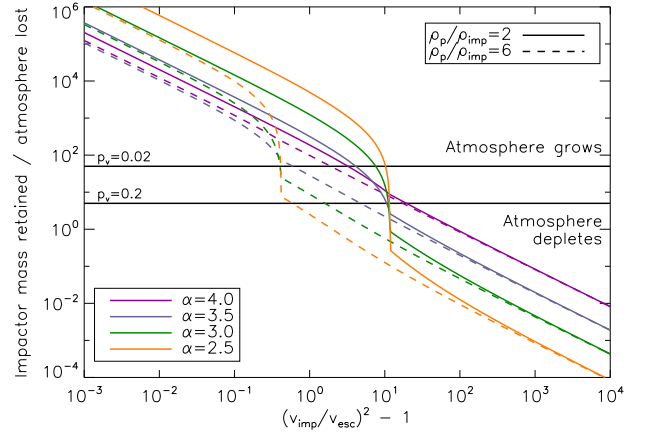}
    \end{tabular}
    \vspace{-2.55in}
    \caption{Ratio of total impactor mass retained to atmospheric mass lost for impacts from a
    size distribution as a function of the ratio of impact velocity to escape velocity $v_{\rm imp}/v_{\rm esc}$.
    The solid and dashed lines are for planet to impactor density ratios of
    $\rho_{\rm p}/\rho_{\rm imp}=2$ and 6 respectively.
    The different colours are for different slopes in the size distribution $\alpha$, which is assumed to
    extend from $\eta_{\rm min}=10^{-3}$ to $\eta_{\rm max}=10^9$.
    All calculations use the prescription in \citet{Shuvalov2009}.
    The growth or depletion of an atmosphere in impacts is determined by whether the plotted ratio is more or less
    than $1/p_{\rm v}$, where $p_{\rm v}$ is the fraction of retained impactor that goes into the atmosphere,
    two representative values for which are shown with horizontal lines.
    }
   \label{fig:fv}
  \end{center}
\end{figure}

Eqs.~\ref{eq:matmlosstot} and \ref{eq:mimpacctot} can be used to determine the ratio of impactor mass retained to that lost
from the atmosphere which is shown in Fig.~\ref{fig:fv} for $\eta_{\rm min}=10^{-3}$ and $\eta_{\rm max}=10^9$.
As long as the range of impactor sizes is large enough for the integrals in these equations to be independent of the boundaries,
the resulting ratio depends only on the ratio of the impact velocity to the planet's escape speed ($v_{\rm imp}/v_{\rm esc}$),
the slope in the size distribution ($\alpha$), and the ratio of planet to impactor densities ($\rho_{\rm p}/\rho_{\rm imp}$).
This shows that whether a planet gains or loses mass (i.e., whether the plotted ratio is more or less than unity), is determined
primarily by the impact velocity in that mass gain requires small $v_{\rm imp}/v_{\rm esc}$.
For size distributions in which the mass is dominated by large impactors (i.e., $\alpha<4$), the ratio shows a sharp
increase for low impact velocities $v_{\rm imp}/v_{\rm esc} < 7.1\rho_{\rm imp}/\rho_{\rm p}$, since this is the
threshold below which all large impactors with $\eta>1000$ can contribute to mass gain (see eq.~\ref{eq:mimpacc}).
The magnitude of the increase is greater for distributions that are more strongly weighted towards large impactors
(i.e., lower $\alpha$), and in this regime the ratio inevitably depends on the assumptions about $\eta_{\rm max}$.
Similar reasons explain why the ratio has a steeper dependence on impact velocity just above this threshold
for size distributions with smaller $\alpha$, in this case because of the increased retention of intermediate-sized impactors.
Mass gain is also favoured for higher impactor densities (i.e., smaller $\rho_{\rm p}/\rho_{\rm imp}$).
The size distribution also plays a role, in that distributions with impactor mass weighted more toward small planetesimals
(i.e., higher $\alpha$) tend to favour mass gain, since all small planetesimals are retained.
However, this trend is reversed (i.e., mass gain is favoured for smaller $\alpha$) for cases where both impactor velocities are
small ($v_{\rm imp}/v_{\rm esc} \ll 1$) and impactor densities are high (i.e., small $\rho_{\rm p}/\rho_{\rm imp}$),
since in this case impactors larger than those that dominate atmospheric mass loss can be retained;
this occurs when $\eta_{\rm maxret} \gg 10^3$ (see eq.~\ref{eq:etamaxret} and Fig.~\ref{fig:mm}), which given that
$v_{\rm imp}/v_{\rm esc} \geq 1$ can only happen for small $\rho_{\rm p}/\rho_{\rm imp}$.
One further consideration is required to determine the effect on the planet's atmosphere, i.e., whether this grows or
depletes with time, which is the fraction of the impactor mass that is retained that goes into the atmosphere $p_{\rm v}$
(see horizontal lines on Fig.~\ref{fig:fv}).

To quantify the effect of the limits of the integration, we determined from eq.~\ref{eq:matmlosstot} the range of $\eta$ above and below
which contributed $10$\% of the total mass loss (and likewise for impactor retention from eq.~\ref{eq:mimpacctot}).
This showed that, as might be expected from Fig.~\ref{fig:mm},
80\% of the atmospheric mass loss originates in a narrow range of $\eta$ that depends only on $\alpha$, which is
from $10^{1.8}-10^{4.2}$ for $\alpha=4$, $10^{2.0}-10^{4.8}$ for $\alpha=3.5$, and $10^{2.8}-10^{7.8}$ for $\alpha=2.5$.
The impactor mass that is retained comes from a larger range of $\eta$ that depends on all variables.
In particular, for $\alpha \geq 4$ the lower limit of $\eta_{\rm min}$ cannot be ignored, because 
all of the mass of impactors smaller than $\eta<10$ are retained and for such size distributions the
mass is weighted towards the smallest impactors (or is equal in logarithmically spaced bins for $\alpha=4$).
As such, Fig.~\ref{fig:fv} is only valid for $\alpha=4$ for the specific case of $\eta_{\rm min}=10^{-3}$ and care is needed
when considering such steep distributions for which impactor retention likely dominates.
For $\alpha=3.5$ the range of $\eta$ contributing to impactor mass retention is better defined, and if $\eta_{\rm min}$ is decreased
to arbitrarily low values, it is found that 80\% of the mass retention comes from a range in $\eta$ of $10^{-4}$ up to
around 10, but could be higher up to $\eta_{\rm maxret}$ from eq.~\ref{eq:etamaxret}.
Since mass retention is weighted to larger $\eta$ when the impact velocity drops below the threshold
of $7.1\rho_{\rm imp}/\rho_{\rm p}$, the $\eta_{\rm max}$ limit becomes an important consideration for such low velocities,
as noted in the previous paragraph.
The situation is similar for $\alpha=2.5$, except that smaller impactors contribute less such that the lower limit is now closer to $10^{-1}$.
These ranges of $\eta$ should be used in conjunction with eq.~\ref{eq:eta} to determine whether a given size range falls inside these
limits.
Thus, the typical range of sizes that contribute to the growth and loss of mass from an Earth-like atmosphere for $\alpha=3.5$
is 0.02-1\,km for growth and 2-20\,km for loss.

\subsection{Effect of multiple impacts on atmosphere evolution}
\label{ss:atmevol}
The results from \S \ref{ss:mult} can now be used to improve on the model of atmospheric evolution from \S \ref{s:simp}.
We will return in \S \ref{ss:fv} to what \S \ref{ss:mult} predicts for the value of $f_{\rm v}$.
For now we note that, for a given scenario, it is reasonable to assume (as was also assumed in \S \ref{s:simp}) that
$f_{\rm v}$ remains constant throughout the evolution.
This is because $f_{\rm v}$ can be determined from the ratio plotted in Fig.~\ref{fig:fv} by multiplying
by the fraction of the impactor mass that is retained that goes into the atmosphere (i.e., $p_{\rm v}$).
The ratio plotted in Fig.~\ref{fig:fv} has already assumed and then averaged over a given size distribution of
impactors ($\alpha$), and assumed an impactor density ($\rho_{\rm imp}$), so for a given scenario the plotted ratio just needs to be
averaged over the distribution of impact velocities.
All of these will depend on the scenario assumed (e.g., the location and mass of the planet, and the provenance
of the impactors), but will not depend on the mass of the atmosphere,
as long as the size distribution is broad enough, and other parameters like impact velocity appropriate, for the limits in the
integrals in equations~\ref{eq:matmlosstot} and \ref{eq:mimpacctot} to be unimportant.
This caveat on the limits of the integrals is important however, since they cannot always be ignored and
\S \ref{ss:airless} considers the situation in which the planet starts with no atmosphere where this is certainly
not possible.

\begin{figure}
  \begin{center}
    \vspace{-0.6in}
    \begin{tabular}{c}
      \hspace{-0.55in} \includegraphics[width=1.25\columnwidth]{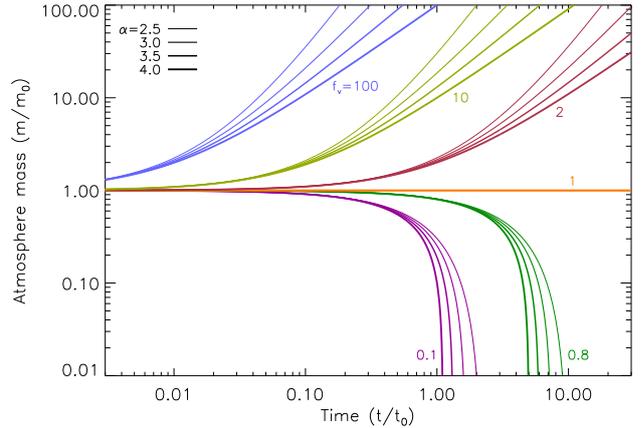}
    \end{tabular}
    \vspace{-2.55in}
    \caption{Updated model from Fig.~\ref{fig:simp} for the evolution of atmospheric mass in which the rates of gain (due to accretion of volatiles
    that replenishes a secondary atmosphere) and loss (that depletes both primordial and secondary atmospheres) both scale with atmosphere mass.
    The evolution depends only on the ratio of the gain and loss rates given by the parameter $f_{\rm v}$ (six different values of which
    are shown with different colours as noted in the annotation) and the slope in the size distribution $\alpha$ (denoted by the different
    thickness lines).
    For clarity only the total atmospheric mass is shown, since the contribution of the primordial and
    secondary components can be inferred from Fig.~\ref{fig:simp} which is identical to that for $\alpha=4$, and is similar for the other values
    of $\alpha$.
    }
   \label{fig:matm2}
  \end{center}
\end{figure}

What \S \ref{ss:mult} does show, however, is that the model of \S \ref{s:simp} can no longer assume that mass loss and
gain are independent of time, since eqs.~\ref{eq:matmlosstot}-\ref{eq:a} show that these should instead be proportional
to $m^{(4-\alpha)/3}$.
This arises because as the atmosphere decreases in mass it is smaller planetesimals that dominate the atmospheric mass loss,
because the larger planetesimals can only remove the atmosphere in the vicinity of the impact \citep[e.g.,][]{Melosh1989};
a similar argument applies as the atmosphere grows.
We implement this into the model by assuming $\dot{m}^{-} = \dot{m}^{-}_{0} (m/m_0)^{(4-\alpha)/3}$ and
$\dot{m}^{+}_{\rm v} = f_{\rm v} \dot{m}^{-}$, where $\dot{m}^{-}_{0}$ is a constant equal to the initial mass loss rate.
This results in the following evolution
\begin{eqnarray}
  m/m_0         & = &  \left[1 + \left(\frac{\alpha-1}{3}\right)(f_{\rm v}-1)(t/t_0)\right]^\frac{3}{\alpha-1}, \label{eq:mm0}
\end{eqnarray}
with $m_{\rm p}$ from eq.~\ref{eq:mpm0}, $m_{\rm v} = m - m_{\rm p}$ and $t_0=m_0/\dot{m}^{-}_{0}$.
This evolution is shown in Fig.~\ref{fig:matm2} for $\alpha=[2.5,3,3.5,4]$, and is the same as that of Fig.~\ref{fig:simp}
for $\alpha=4$ (since this results in mass loss that is independent of atmospheric mass), noting however that the
model is invalid for size distributions with $\alpha \geq 4$ because in this case (as noted in \S \ref{ss:mult}) the
lower limit $\eta_{\rm min}$ becomes important in the calculation of $f_{\rm v}$, which thus varies with time.

Fig.~\ref{fig:matm2} shows that the evolution is not much different with this change.
The atmosphere still disappears in a finite time for $f_{\rm v}<1$ and grows monotonically with time for $f_{\rm v}>1$ and $t/t_0 \gg 1$.
The timescale on which the evolution takes place now depends on the slope in the size distribution, with shallower size
distributions (i.e., smaller $\alpha$, meaning more weighted to large impactors) resulting in atmospheres being lost
more slowly or growing more rapidly.
However, the sense of faster or slower here is in units of dimensionless time which is itself dependent on $\alpha$ through
the initial mass loss rate, and so it is not possible from this alone to determine whether the evolution takes more or less
real time.
Eq.~\ref{eq:mm0} shows that the time for the atmosphere to be completely lost for $f_{\rm v}<1$ is
\begin{equation}
  t_{\rm bare}=\left( \frac{3}{\alpha-1} \right) \left( \frac{1}{1-f_{\rm v}} \right) \left( \frac{m_0}{\dot{m}^{-}_0} \right).
  \label{eq:tbare}
\end{equation}
This means that an $f_{\rm v}<1$ planet must accrete a total impactor mass of
\begin{equation}
  \Delta m_{\rm ac,bare} = m_0 \left( \frac{m_{\rm ac}}{m_{\rm atmloss}} \right)_0 \left( \frac{3}{\alpha-1} \right) \left( \frac{1}{1-f_{\rm v}} \right)
  \label{eq:macbare}
\end{equation}
to completely lose its atmosphere, where $(m_{\rm ac}/m_{\rm atmloss})_0$ is the inverse of the ratio from eq.~\ref{eq:matmlosstot} calculated for
the initial atmosphere.
This is similar to the mass required to double the atmosphere in the case that $f_{\rm v}>1$,
which is $[2^{(\alpha-1)/3}-1]\Delta m_{\rm ac,bare}$.

To summarise, Fig.~\ref{fig:matm2} can be used to determine the effect of multiple impacts on a planet's atmosphere.
This requires calculation of $f_{\rm v}$ which must be done from Fig.~\ref{fig:fv} as discussed in \S \ref{ss:fv}.
Such calculation is complicated by the fact that the plotted curves need to be averaged over
the appropriate distribution of impact velocities and impactor densities, and an assumption
needs to be made about the fraction of the impactor mass that is retained that goes into the atmosphere ($p_{\rm v}$).
There are also a few caveats.
First, this assumes that the calculations that go into Fig.~\ref{fig:fv}
are not affected by the largest or smallest impactors in the distribution. 
Also, this assumes that the evolution in a given timestep can be well described by the average mass loss, which thus
ignores the possible stochastic contribution of single giant impacts (see \S \ref{ss:stochasticity}).
Finally, an increase with time of the volatile content of the planet's atmosphere would increase its mean molecular weight $\mu$.
While this would have no effect on $f_{\rm v}$, and so whether the atmosphere would ultimately grow or deplete, this would affect the
evolutionary timescale which would get longer as the atmosphere gets more volatile-rich.
This is because of the reduced atmospheric scale height (eq.~\ref{eq:h}) which results in a decreased mass change per
colliding mass (eqs.~\ref{eq:matmlosstot}-\ref{eq:a}).
Some of these complications and caveats will be explored further in \S \ref{ss:fv} after which the particular case of
the evolution of a planet that starts without an atmosphere will be discussed in \S \ref{ss:airless}.

\subsection{Determining $f_{\rm v}$}
\label{ss:fv}
As discussed in \S \ref{ss:atmevol}, calculation of $f_{\rm v}$ can be done from Fig.~\ref{fig:fv}
by averaging over the appropriate distribution of impact velocities and impactor densities,
making also an assumption about the fraction of the impactor mass that is retained that
goes into the atmosphere ($p_{\rm v}$).
The further assumptions about impactor types used in this paper are discussed in
\S \ref{sss:imptypes} before using these in \S \ref{sss:fv} to determine $f_{\rm v}$
for planets in different regions of parameter space, and considering the sensitivity of the
derived $f_{\rm v}$ to the assumptions in \S \ref{sss:fvchanges}.

\subsubsection{Assumptions about impactor types}
\label{sss:imptypes}
Assumptions in the literature about both impactor densities and the impactor mass retained typically involve an assumption about
whether the impacting body is asteroidal or cometary.
While this terminology refers to Solar system-like objects, we will apply this more generally here
with the following meaning.
We will assume asteroidal impactors to have a density of $\rho_{\rm impa}=2.8$\,g\,cm$^{-3}$ and that $p_{\rm va}=2$\% of their
mass goes into the atmosphere on impact, which is based on this being the approximate volatile content
of carbonaceous chondrites \citep[e.g.,][]{Grady2003, Sephton2002}
excluding water which might precipitate onto the surface for planets in the habitable zone \citep[e.g.,][]{Zahnle2007}.
These volatiles would be in the form of insoluble organic macromolecular material, soluble organics and carbonates,
and may be expected to be degassed during impacts leading to atmospheres rich in H$_2$O, H$_2$, CO or CO$_2$
\citep[e.g.,][]{Schaefer2010}.
Cometary impactors will be assumed to have a density $\rho_{\rm impc}=0.9$\,g\,cm$^{-3}$ with $p_{\rm vc}=20$\% of their mass
going into the atmosphere on impact for similar reasons, with the majority of the volatiles in the form of CO, CO$_2$ and O$_2$
(excluding water again for the same reason as for asteroidal impactors),
and a smaller fraction in molecules such as methane, ethane, methanol, formaldehyde, ammonia, hydrogen cyanide, hydrogen sulfide
\citep[e.g.,][]{Mumma2011, Rubin2019}.
These assumptions should serve to indicate outcomes for two different types of impactor, but are not suppposed to
represent the only possible impactor types.

The distribution of impactor velocities is usually taken from N-body simulations of impactor populations as they
interact with a planetary system.
Since such simulations require an assumption about the source of the impactors and the planetary system that results
in them evolving onto orbits that can result in a collision with the planet in question, we prefer to avoid detailed
simulations here.
Rather we base the expected range of impactor velocities on the following analytical considerations \citep[see also][]{Kral2018}.
Consider a planet of mass $M_{\rm p}$ on a circular orbit at $a_{\rm p}$ interacting with an
impactor on a comet-like orbit, which here we take to mean one with an eccentricity that is close to 1.
The impactor's orbital velocity at the location of the planet is approximately $\sqrt{2}v_{\rm p}$, where
$v_{\rm p}=\sqrt{GM_\star/a_{\rm p}}$ is the orbital velocity of the planet.
If the inclination of the impactor's orbit relative to that of the planet is small then their relative velocity on approach
to impact is $[3-2\sqrt{2q/a_{\rm{p}}}]^{1/2}v_{\rm p}$, where $q$ is the impactor's pericentre distance.
This relative velocity is thus in the range $(\sqrt{2}-1)v_{\rm p}$ (if the comet is close to pericentre at impact) to
$\sqrt{3}v_{\rm p}$ (if the comet's pericentre is far inside the planet's orbit), i.e., $(0.4-1.7)v_{\rm p}$.
Impactors that originated in an asteroid belt or indeed from the vicinity of the planet in question
may have a lower relative velocity at impact, of order $\sqrt{1.5}ev_{\rm p}$ for distributions with mean
eccentricity $e$ and mean inclination $e/2$ \citep{Wetherill1993}.

While impact velocities might be expected to come from a distribution,
we take one value as being representative for the resulting $f_{\rm v}$, which could be derived for a given distribution of impact
velocities by implementing this in eqs~\ref{eq:matmlosstot} and \ref{eq:mimpacctot} and then averaging the resulting ratio.
Here we assume the relative velocities are $\xi v_{\rm p}$, where $\xi_{\rm c}=1.0$ for cometary impactors and $\xi_{\rm a}=0.3$ for 
asteroidal impactors, then account for the effect of gravitational focussing to get for impact velocities
\begin{eqnarray}
  v_{\rm imp}/v_{\rm esc} & = & \sqrt{1+(\xi v_{\rm p}/v_{\rm esc})^2}, \label{eq:vimpvesc} \\
  v_{\rm p}/v_{\rm esc}   & = & 3.4M_\star^{1/2}a_{\rm p}^{-1/2}M_{\rm p}^{-1/3}\rho_{\rm p}^{-1/6}, \label{eq:vpvesc}
\end{eqnarray}
for the units in Table~\ref{tab:units}.
It is worth re-iterating that N-body simulations are needed to get an accurate distribution of $\xi$ if the dynamical origin
of the impactors is known.
For example, our assumed values are slightly more extreme than those which might be inferred for asteroids and comets impacting the Earth
during the Late Heavy Bombardment; e.g., figs. 6 and 7 of \citet{deNiem2012} suggest (by eye) average values closer to $\xi_{\rm a}=0.5$ and
$\xi_{\rm c}=0.8$.
Similarly, fig.~7 of \citet{Kral2018} shows that the distribution of impact velocities for planets in a chain can depend
on the location in that chain, while our simplistic approach overestimates by a factor of 2 the median impact velocity
for the outermost planets in the TRAPPIST-1 system (f, g and h), and underestimates it for the innermost planets.
Such details may contribute to any differences in our results to studies using N-body simulations, but this should not affect
general trends, and this can be accounted for where N-body simulations are available.

\subsubsection{$f_{\rm v}$ for different planets}
\label{sss:fv}

\begin{figure*}
  \begin{center}
    \vspace{-0.05in}
    \begin{tabular}{cc}
      \hspace{-0.1in} Asteroidal Impactors & \hspace{-0.5in} Cometary Impactors \\[-0.6in]
      \hspace{-0.45in} \includegraphics[width=1.25\columnwidth]{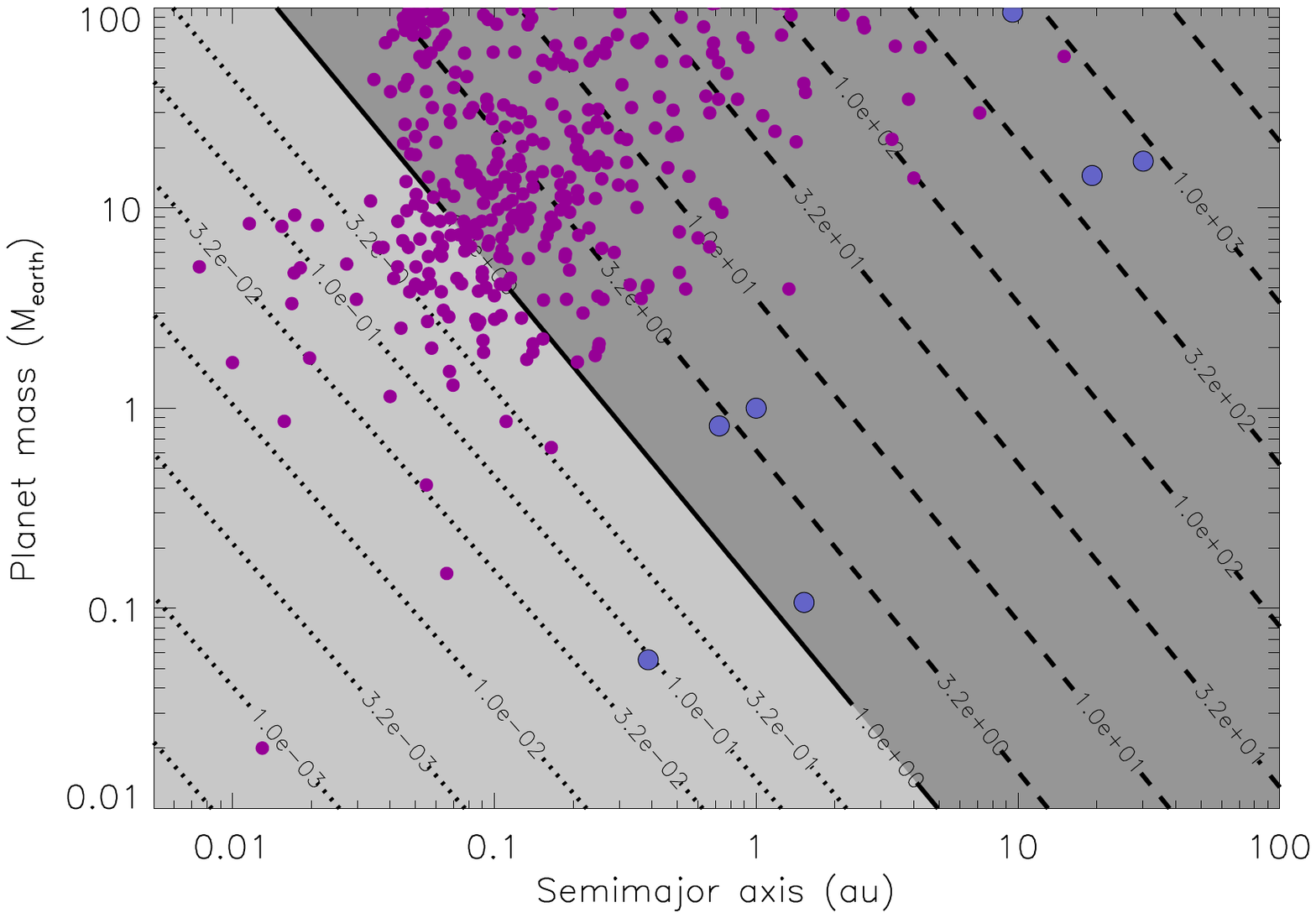} &
      \hspace{-0.85in} \includegraphics[width=1.25\columnwidth]{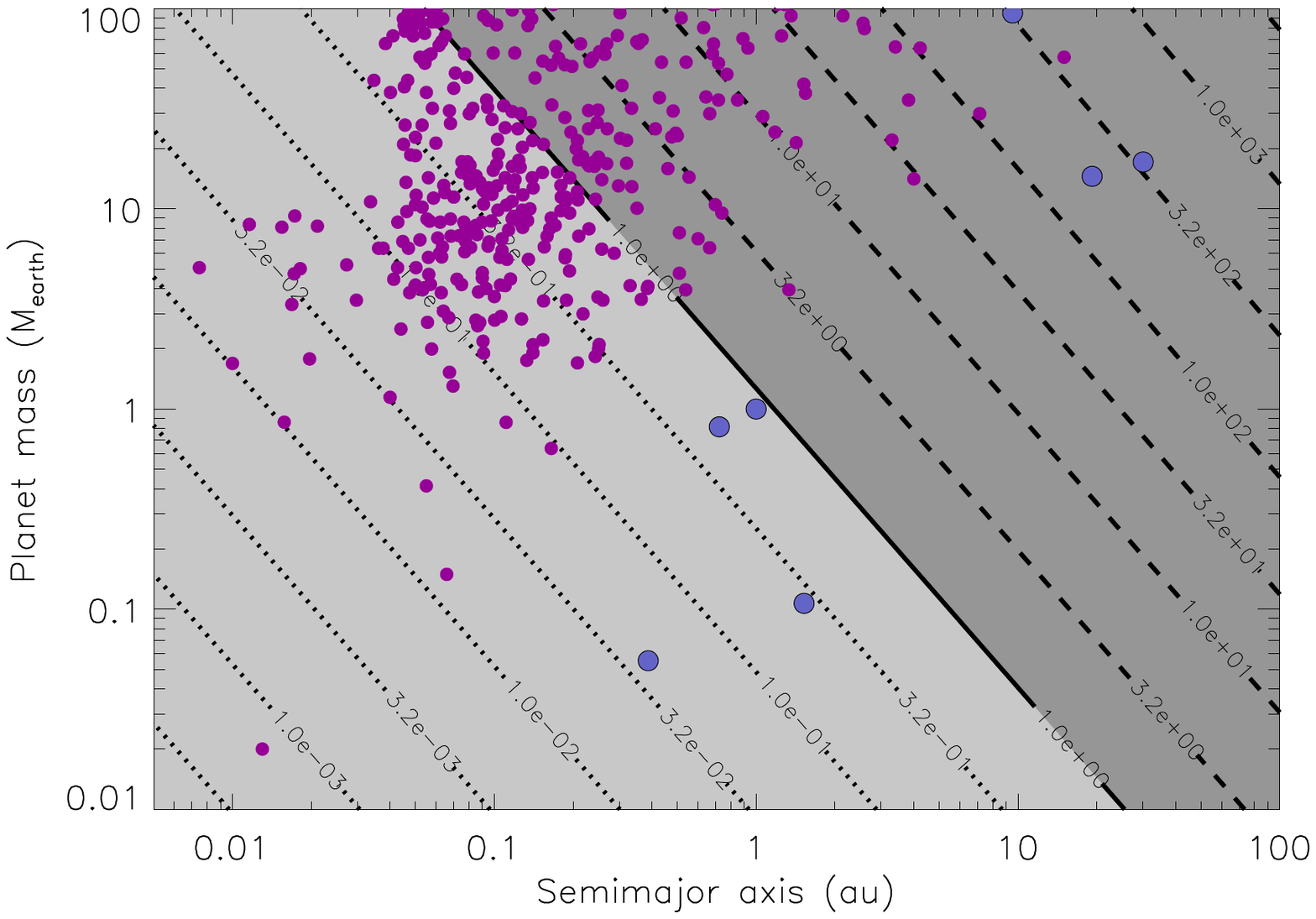} \\[-3.05in]
      \hspace{-0.45in} \includegraphics[width=1.25\columnwidth]{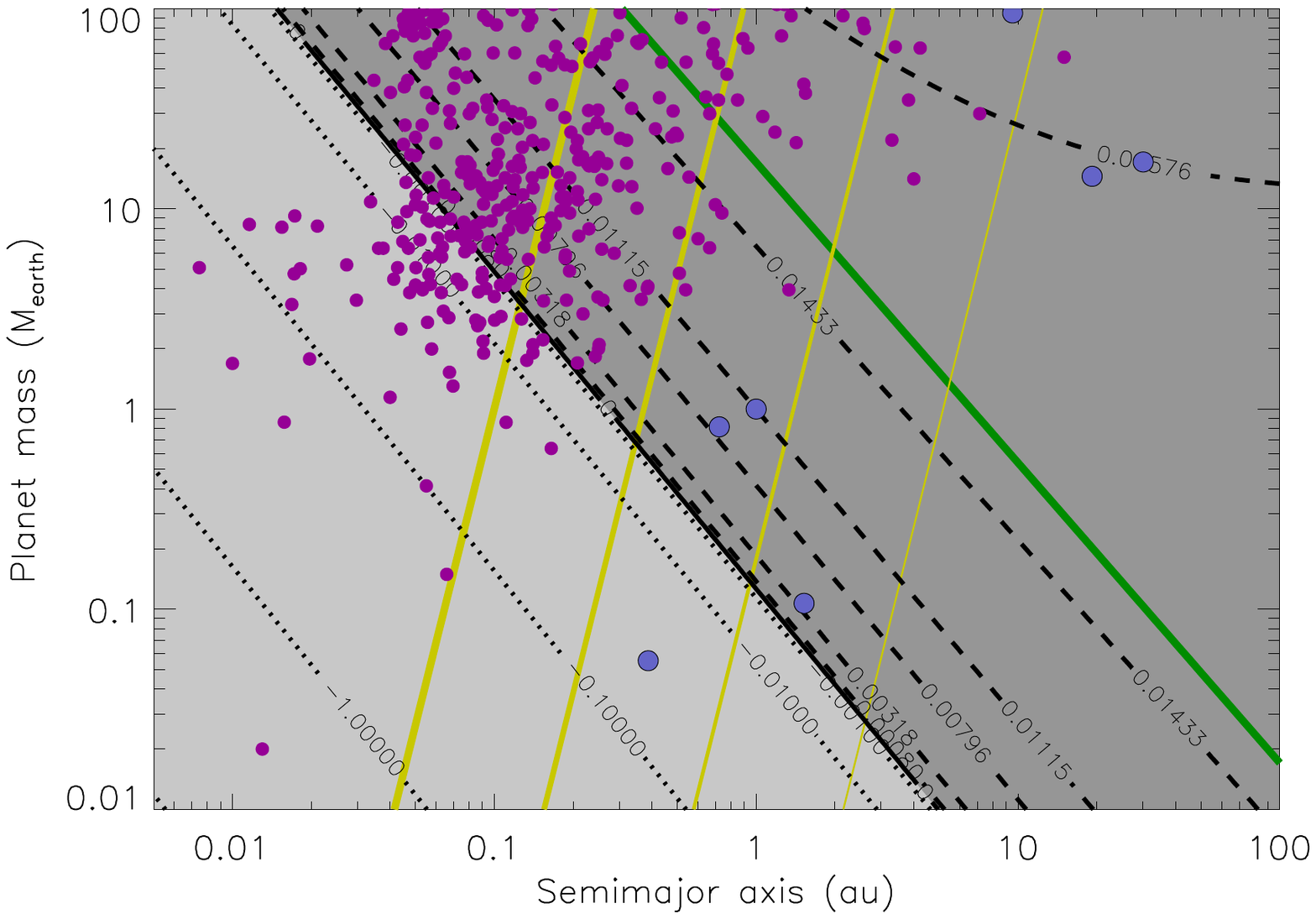} &
      \hspace{-0.85in} \includegraphics[width=1.25\columnwidth]{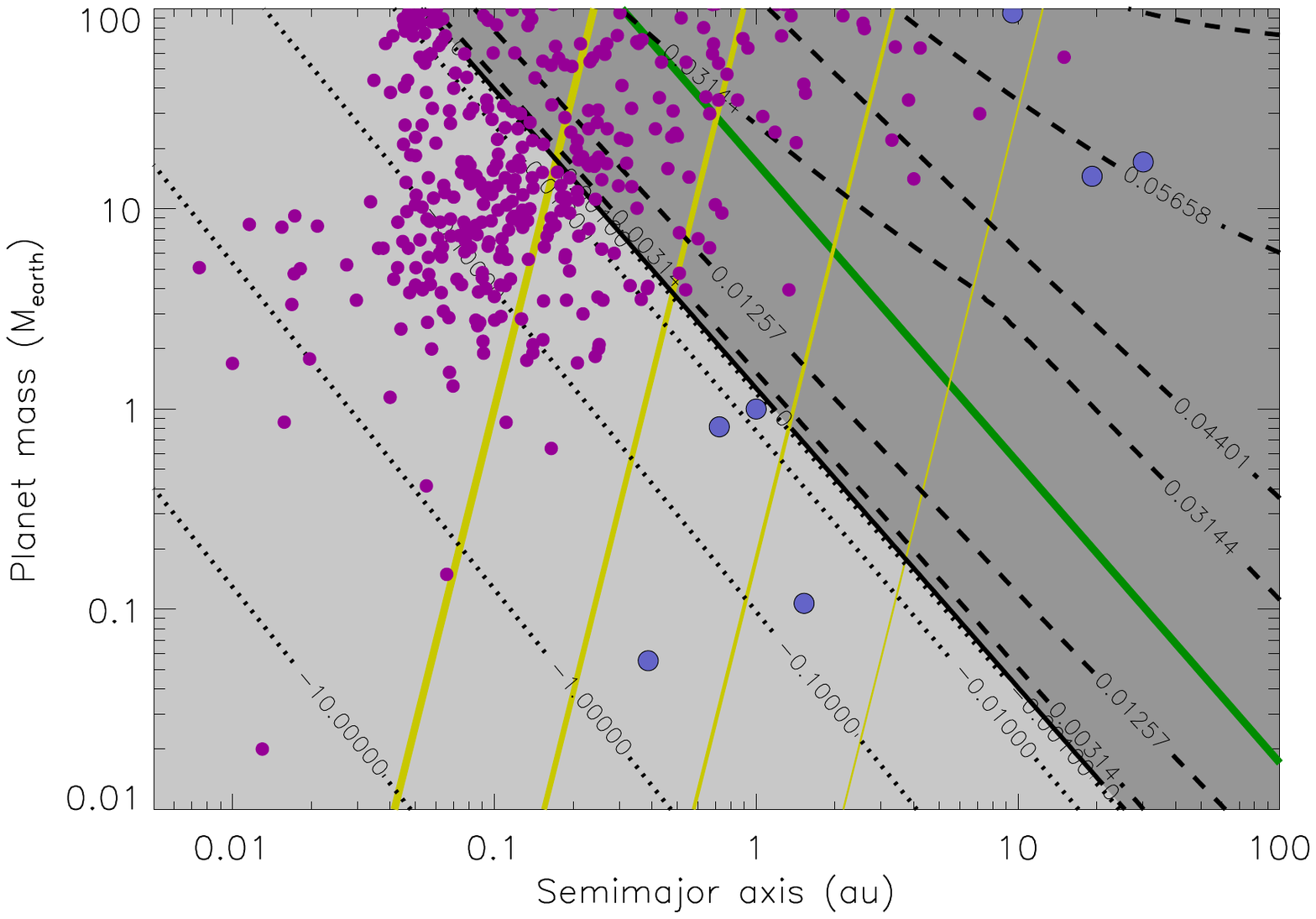} \\[-3.05in] 
      \hspace{-0.45in} \includegraphics[width=1.25\columnwidth]{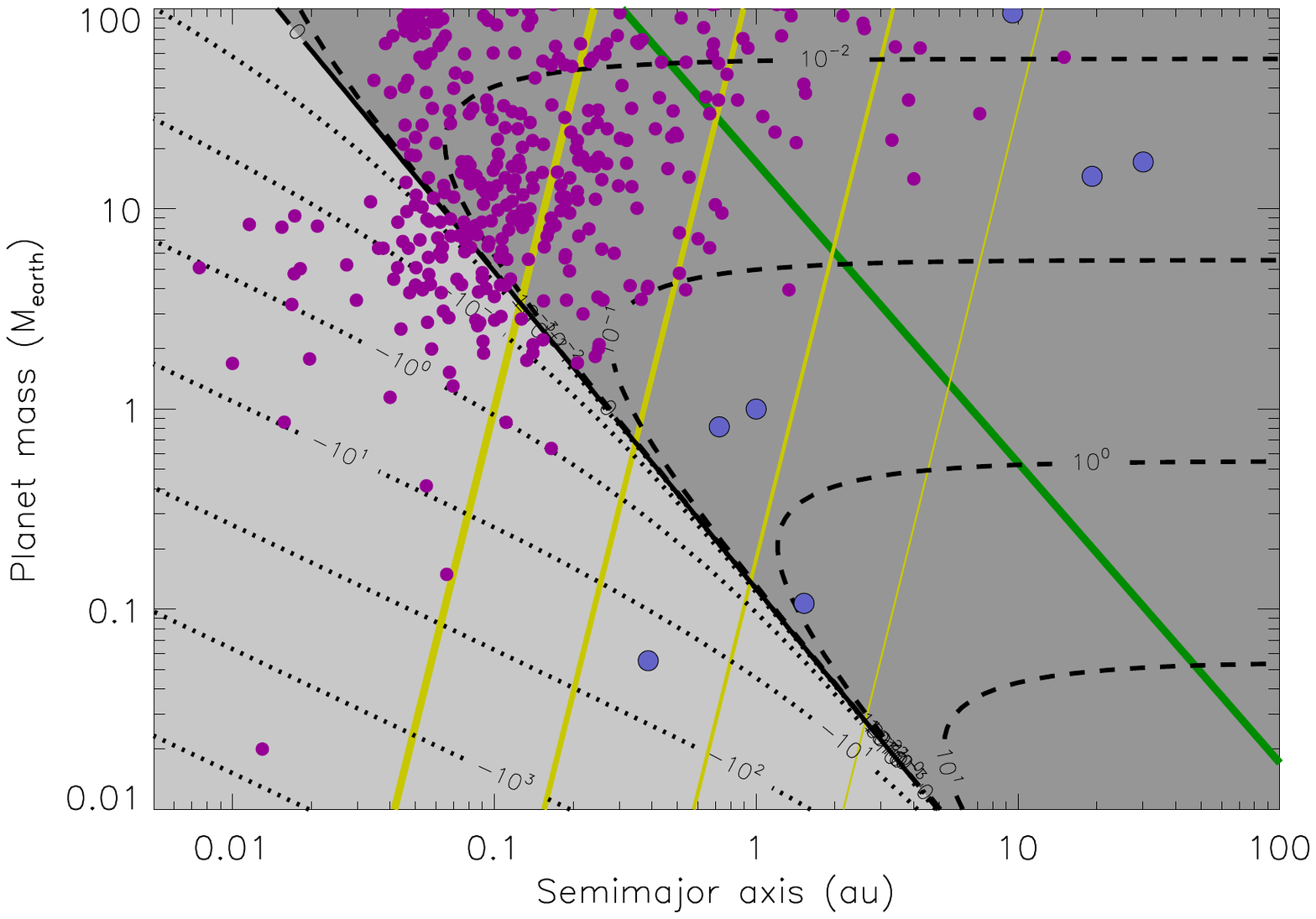} &
      \hspace{-0.85in} \includegraphics[width=1.25\columnwidth]{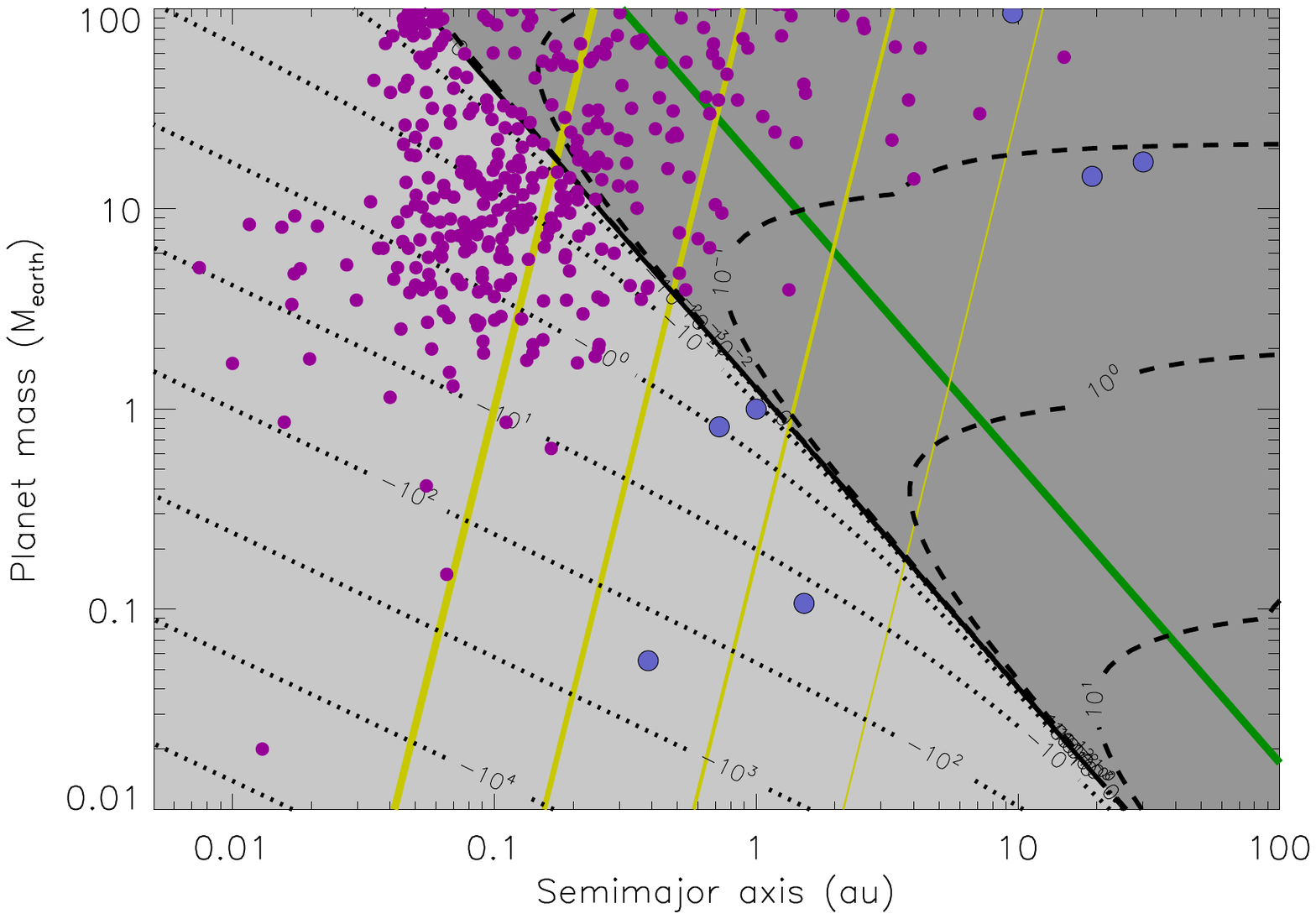} 
    \end{tabular}
    \vspace{-2.55in}
    \caption{Outcome of impacts with planets of different masses and semimajor axes orbiting $1{\rm M}_\odot$ stars.
    The left column assumes asteroidal impactors ($\rho_{\rm imp}=2.8$\,g\,cm$^{-3}$ with 2\% volatiles and relative velocities approaching impact
    of 0.3 times the planet's orbital velocity),
    while the right column assumes cometary impactors ($\rho_{\rm imp}=0.9$\,g\,cm$^{-3}$ with 20\% volatiles and relative velocities approaching impact
    of 1.0 times the planet's orbital velocity).
    In all panels an impactor size distribution with $\alpha=3.5$ from $D_{\rm min}=1$\,m to $D_{\rm max}=100$\,km
    is assumed, and the planet is assumed to have a density 5.5\,g\,cm$^{-3}$, and a $\mu=29$ atmosphere with a
    mass $0.85 \times 10^{-6}$ that of the planet.
    In the top row contours show the ratio of atmospheric mass gain (due to volatile retention)
    to mass loss (due to atmosphere stripping) in planetesimal impacts, i.e., $f_{\rm v}$.
    In the middle row contours show the change in atmosphere mass per accreted impactor mass,
    i.e., $\Delta m/\Delta m_{\rm ac}$.
    In the bottom row contours show the fractional change in atmosphere mass after accreting
    $m_{\rm ac}=3 \times 10^{-5}{\rm M}_\oplus$.
    The solid black line is the impact shoreline; 
    the atmospheres of planets above this line (i.e., in the darker shaded region where contours are dashed) gain mass in collisions,
    while those below (i.e., in the lighter shaded region where contours are dotted) lose mass.
    The dark green line is that for $v_{\rm esc}/v_{\rm p}=1$ above which the planet is more likely to eject planetesimals it interacts with
    than be impacted by them.
    The lighter green lines are for constant accretion timescale from a comet-like population, where that timescale for lines from left to
    right (from thicker to thinner lines) is 0.3\,Myr, 30\,Myr, 3\,Gyr and 300\,Gyr.
    The accretion efficiency is reduced for planets with longer collision timescales, since it is more likely that other processes remove planetesimals
    from the vicinity of the planet before impacts occur.
    The purple circles are known exoplanets for $0.6-1.4{\rm M}_\odot$ stars \citep[from the exoplanet.eu database on 28 November 2018,][]{Schneider2011}.
    The larger blue circles are the Solar system planets.
    }
   \label{fig:outcome1}
  \end{center}
\end{figure*}

We can now determine for our assumptions about asteroidal or cometary impactors what $f_{\rm v}$ is for planets with
different masses, semimajor axes and densities, with additional free parameters of the stellar mass and the slope in
the size distribution of impactors.
The top panels of Fig.~\ref{fig:outcome1} show the resulting $f_{\rm v}$ for planets of density
5.5\,g\,cm$^{-3}$ (i.e., Earth-like) orbiting solar mass stars for the two different impactor types
assuming an impactor distribution with $\alpha=3.5$ between $D_{\rm min}=1$\,m and $D_{\rm max}=100$\,km
and an atmosphere mass $\delta=0.85 \times 10^{-6}$ times that of the planet mass
(which means it is Earth-like in terms of its relative mass, but not necessarily in terms of its
surface pressure, see eq.~\ref{eq:p}).
For reference the locations of known exoplanets\footnote{Taken on 28 November 2018 from the exoplanet.eu database \citep{Schneider2011}.}
and the Solar system planets are also shown.
The slope in the contours of equal $f_{\rm v}$ arises because this ratio is the same for planets with the same ratio of
escape velocity to orbital velocity, which is for planets for which $M_{\rm p} \propto a_{\rm p}^{-3/2}$
(see eq.~\ref{eq:vpvesc}).
This essentially shows the susceptibility of planets in different regions of parameter space to erosion or growth by planetesimal
impacts, since as noted in \S \ref{ss:atmevol}, this determines whether the atmospheres grow or deplete given sufficient impacts.
The $f_{\rm v}=1$ division between the different outcomes we call the impact shoreline, by analogy with the cosmic shoreline
discussed in \citet{Zahnle2017}.

Comparison of asteroidal and cometary impactors (left and right panels on Fig.~\ref{fig:outcome1}) shows that planets
are more susceptible to mass loss for impactors with the assumed cometary properties, because the additional volatile content of such impactors
is not sufficient to offset the destructiveness of their greater impact velocity.
Thus, for the given assumptions, the Earth's atmosphere and that of Venus would be expected to grow in collisions with asteroids, but to
deplete in collisions with comets, while all impacts would deplete the atmospheres of Mars and Mercury.
For the given assumptions, the atmospheres of many of the known exoplanets would be predicted
to grow in all types of planetesimal impacts.
This means that, should they have undergone significant bombardment (which will be quantified in the next sections), their
atmospheres may be more massive or more volatile-rich compared to their primordial values.
However, planets that are close enough to the star, in particular those that underwent bombardment
by comet-like impactors, would have had their atmospheres stripped.

\subsubsection{How $f_{\rm v}$ changes with different assumptions}
\label{sss:fvchanges}
\begin{figure}
  \begin{center}
    \vspace{-0.6in}
    \begin{tabular}{c}
      \hspace{-0.55in} \includegraphics[width=1.25\columnwidth]{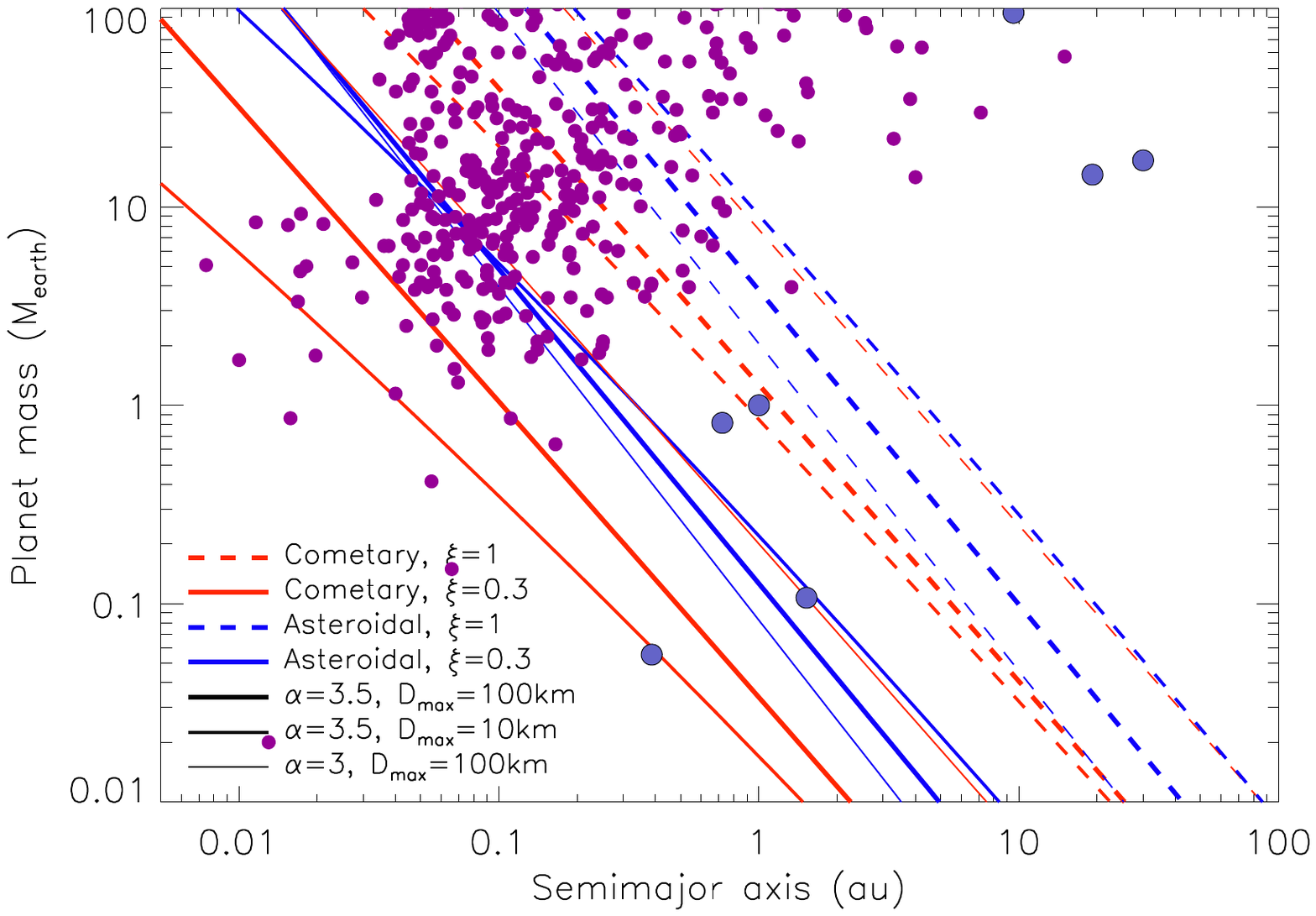}
    \end{tabular}
    \vspace{-2.55in}
    \caption{The dependence of the impact shoreline on impactor assumptions. 
    The lines delineate between atmospheres that grow (upper right) and deplete (bottom left) in impacts for 
    planets orbiting Sun-like stars.
    Different assumptions about the impactors are shown with different lines.
    Asteroidal impactors ($\rho_{\rm imp}=2.8$\,g\,cm$^{-3}$ with 2\% volatiles) are shown with blue lines and
    cometary impactors ($\rho_{\rm imp}=0.9$\,g\,cm$^{-3}$ with 20\% volatiles) with red lines.
    Solid lines are for relative velocities near impact 0.3 times the planet's orbital velocity, while dashed lines have those relative velocities
    equal to the planet's orbital velocity.
    Lines of different thickness indicate different assumptions about the slope in the impactor size distribution ($\alpha$) and maximum
    impactor size ($D_{\rm max}$) as shown in the legend (in all cases $D_{\rm min}=1$\,m is assumed).
    For all lines the planet is assumed to have a density 5.5\,g\,cm$^{-3}$, and a $\mu=29$ atmosphere with a
    mass $0.85 \times 10^{-6}$ that of the planet.
    }
   \label{fig:fvch}
  \end{center}
\end{figure}

While a specific atmosphere mass and upper and lower limits to impactor size were assumed when making the
top panels in Fig.~\ref{fig:outcome1}, for the reasons given in \S \ref{ss:mult} these should have little effect
on the resulting calculation of $f_{\rm v}$ in the sense that the outcome would have been very similar with
different atmosphere masses (if not too different, see \S \ref{ss:airless}) and with the assumption
that the size distribution had extended to arbitrarily large and small values.
A finite upper or lower limit to impactor sizes can become important, however, in certain circumstances.
For example, given the dominating impactor sizes noted at the end of \S \ref{ss:mult} for the Earth
(i.e., 0.02-1\,km for impactor retention and 2-20\,km for atmosphere loss), an upper limit on planetesimal size
in the 1-10\,km range would have the effect of reducing atmosphere loss without affecting its gain resulting
in an increase in $f_{\rm v}$.
Also, Fig.~\ref{fig:fv} shows that flatter size distributions
(i.e., smaller $\alpha$, weighted more to larger impactors) would result
in more disruptive impacts and so a lower $f_{\rm v}$.
These expectations are confirmed in Fig.~\ref{fig:fvch} which shows the planet for which $f_{\rm v}=1$ (i.e., the transition
between atmosphere growth and depletion in impacts, or the impact shoreline)
for different assumptions about the size distribution with lines of different thickness.
That is the $f_{\rm v}=1$ lines move down when $D_{\rm max}$ is decreased (as impacts become less destructive) and up when $\alpha$ is decreased
(as impacts become more destructive).
	
Fig.~\ref{fig:fvch} also shows how the lines of $f_{\rm v}=1$ change with the assumptions about the impact velocities and
impactor composition.
For example, the lines move up as impact velocities are increased from $\xi=0.3$ to 1.0, because the impacts become more destructive
(see Fig.~\ref{fig:fv}), and impacts tend to favour atmosphere growth (the lines move down) as the
fraction of volatiles contained in the impactor ($p_{\rm v}$) is increased,
though impactor density also plays a role in the plotted values (see Fig.~\ref{fig:fv}).
Overall, one point to take away from Fig.~\ref{fig:fvch} is that the outcome of collisions (i.e., whether atmospheres
grow or deplete in impacts) is sensitive to what is assumed about the impactors, particularly about their impact velocities,
but also about their volatile content, and to a lesser extent their size distribution (although the change on Fig.~\ref{fig:fvch} would have
been more significant for $D_{\rm max}=1$\,km).
Thus any definitive claims about atmosphere evolution require these parameters to be well constrained, which is challenging even
in the Solar system.

As noted above, much of the spread in the lines on Fig.~\ref{fig:fvch} can be understood purely from Fig.~\ref{fig:fv}.
The one parameter that requires further thought is the upper impactor size $D_{\rm max}$, the consequence of
which can be understood by rearranging eq.~\ref{eq:eta}, including also the factor from eq.~\ref{eq:vimpvesc},
to find that the size corresponding to a given $\eta$ is
\begin{equation}
  D \propto \eta^{1/3} m^{1/3} \mu^{-2/3} \xi^{-2/3} M_{\rm p}^{-2/9}
            \rho_{\rm p}^{-4/9} M_\star^{-1/3} L_\star^{1/6}
            (1+\rho_{\rm p}/\rho_{\rm imp})^{1/3}.
  \label{eq:deta}
\end{equation}
This allows to determine how the dominating impactor sizes recalled above for the Earth
(i.e., 0.02-1\,km for impactor retention and 2-20\,km for atmosphere loss)
change with different assumptions,
and so whether this calculation is affected by the impactor size limits.
Equation~\ref{eq:deta} shows that the stellar properties do not play a strong role in how planetesimal size maps onto $\eta$
(e.g., for the same $\eta$ for impacts in the TRAPPIST-1 system as for the Solar system,
the impactor size is reduced by only 65\%),
and neither do planet properties
(e.g., a factor 100 increase in planet mass results in a factor 3 decrease in impactor size for the
same $\eta$, or less if atmosphere mass scales with planet mass),
and neither does the impactor type
(e.g., asteroidal impactors are roughly twice the size as cometary impactors for the same $\eta$).
However, the dependencies on $m$ and $\mu$ mean that the dominating impactors
are 100 times larger than found for the Earth for an atmosphere with $\delta=1$\% of the
mass of the Earth and solar composition.
This means that more massive atmospheres are more susceptible to growth and that, if the upper size cut-off is in
a regime where this becomes important, the lines would move down on Fig.~\ref{fig:fvch}
(since an upper cut-off would then cause a lack of destructive impactors).
This would also be the case for a more primordial atmosphere, which conversely means that the increasing volatile fraction
of a growing atmosphere could make impactors more harmful potentially stalling its growth.

\subsection{Evolution of an atmosphere-less planet}
\label{ss:airless}
One situation in which a planet's atmosphere evolution cannot be considered in the manner described in \S \ref{ss:atmevol} is that in which
the planet starts without an atmosphere, i.e., $m_0=0$.
This is a situation in which the limits of the integrals cannot be ignored, since for the smallest and largest impactors alike $\eta \to \infty$
(eq.~\ref{eq:eta}).
To determine what happens in this case we first consider whether impacts are able to leave any mass in the atmosphere.
For low impact velocities, $v_{\rm imp}/v_{\rm esc}<7.1\rho_{\rm imp}/\rho_{\rm p}$, no mass is retained and so no atmosphere growth is possible
and the planet will remain forever atmosphere-less.

For impact velocities above this limit atmosphere growth will be possible, since $f_{\rm v} \to \infty$, at least initially.
While the atmosphere mass remains small, $\eta_{\rm min}$ will be large (this could mean, e.g., that $\eta_{\rm min} \gg 10^6$),
which would mean from Fig.~\ref{fig:mm} that mass gain exceeds mass loss for all impactor sizes and so $f_{\rm v}$ must be
greater than unity.
Thus the atmosphere would grow with continued bombardment.
As the mass of the atmosphere increases, $\eta_{\rm min}$ (and $\eta_{\rm max}$) would decrease, and the atmospheric mass lost
per impactor mass also grows (as there is more atmosphere to lose) with little change in the mass gain per impactor.
This causes $f_{\rm v}$ to decrease from its initially high value.
Eventually the atmosphere will have grown such that $\eta_{\rm min}$ is small and irrelevant, at which point $f_{\rm v}$ may be
greater than or less than unity.
There may be turning points in the value of $f_{\rm v}$ as a function of atmosphere mass.
If $f_{\rm v}$ remains above unity throughout then the atmosphere will continue to grow indefinitely.
If $f_{\rm v}$ drops below unity then atmosphere growth will stall at the value where $f_{\rm v}$ first reaches unity,
since if it grew further then $f_{\rm v}$ would be less than unity and further impacts
would cause atmosphere loss until $f_{\rm v}$ had increased to unity again
(i.e., $f_{\rm v}=1$ is a stable equilibrium point if $df_{\rm v}/d\delta<0$ at this point).

\begin{figure}
  \begin{center}
    \vspace{-0.6in}
    \begin{tabular}{c}
      \hspace{-0.55in} \includegraphics[width=1.25\columnwidth]{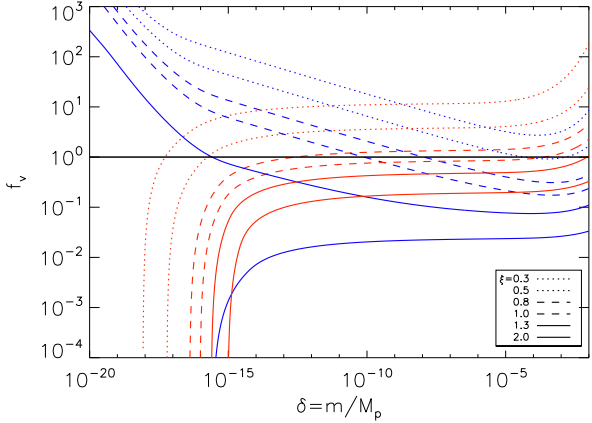}
    \end{tabular}
    \vspace{-2.55in}
    \caption{Dependence of $f_{\rm v}$ on atmosphere mass for an Earth-like planet ($1M_\oplus$, 5.5\,g\,cm$^{-3}$,
    1\,au, $\mu=29$) orbiting a Sun-like star, being impacted by 1\,m-100\,km planetesimals with a size distribution
    $\alpha=3.5$.
    Impactor compositions are assumed to be either
    asteroidal ($p_{\rm v}=0.02$, $\rho_{\rm imp}=2.8$\,g\,cm$^{-3}$, blue lines) or 
    cometary ($p_{\rm v}=0.2$, $\rho_{\rm imp}=0.9$\,g\,cm$^{-3}$, red lines).
    The legend gives the assumed impact velocity in terms of $\xi$.
    }
   \label{fig:fvvm}
  \end{center}
\end{figure}

To illustrate this, Fig.~\ref{fig:fvvm} shows how $f_{\rm v}$ depends on atmosphere mass for the Earth being impacted by
planetesimals of asteroidal and cometary composition at different velocities.
For bombardment by cometary compositions the velocities plotted are all above the transition (which occurs at $\xi=0.24$) and so a bare
Earth would always remain as such.
For asteroidal compositions the transition is at $\xi=1.37$, so for velocities lower than this the atmosphere would grow.
For $\xi=0.5-1.37$ the atmosphere would stall (e.g., at $\delta \approx 10^{-10}$ for $\xi=1.0$), whereas for $\xi<0.5$ the atmosphere
would continue to grow indefinitely.

It is possible to find a combination of impactor parameters that leads to atmosphere growth that stalls at $\delta_\oplus$.
However, before reading too much into Fig.~\ref{fig:fvvm}, a number of uncertainties should be noted.
For example, this prediction depends strongly on the assumptions about the outcome of impacts in the airless limit.
Comparison with other prescriptions \citep[e.g.,][]{Cataldi2017} and simulations \citep[e.g.,][]{Zhu2019} in this limit shows
that the \citet{Shuvalov2009} prescription we are using is reasonable, but may not capture all of the relevant detail.
Also, the atmosphere masses in question are incredibly small, and so the delivery of a single large impactor can be significant;
i.e., the evolution of $\delta$ may be stochastic rather than monotonic at the levels of interest. 
It is also worth noting that similar arguments apply to atmospheres that were predicted in \S \ref{ss:fv} to deplete in impacts,
since if $f_{\rm v}$ increases as the atmosphere depletes (which is necessarily the case for sufficiently low velocities), then
these atmospheres would not be completely removed but instead stall at the value for which $f_{\rm v}$ first goes above unity.
In any case, one thing to take away from Fig.~\ref{fig:fvvm} is that while $f_{\rm v}$ does have some dependence on atmosphere mass,
and one that is particularly important to consider for very low atmosphere masses, it is also relatively flat over a large
range of $\delta$, and so the broad conclusions of previous sections are still valid.

\subsection{Fractional change in atmosphere per cumulative accreted impactor mass}
\label{ss:time}
While \S \ref{ss:fv} considered the susceptibility of a planet's atmosphere to erosion or growth, such
susceptibility does not mean that the atmosphere will completely disappear or grow significantly, as that
requires a consideration of the total mass of impacting planetesimals, their effect on the atmosphere, and
how that compares with the initial atmospheric mass $m_0$.
Clearly, these are not factors that are well known even in the Solar system.
We can however give the reader a feeling for how such considerations may apply to planets in different regions of
parameter space by plotting the model predictions for the ratio of the change in a planet's atmosphere mass to the
mass of impactors accreted, i.e.,
\begin{equation}
  \Delta m/\Delta m_{\rm ac} =  (m_{\rm atmloss}/m_{\rm ac}) (f_{\rm v}-1),
  \label{eq:dmmdmac}
\end{equation}
which is shown in the middle panels of Fig.~\ref{fig:outcome1}.
To make these panels, the assumptions about the initial atmosphere mass (i.e., that this was a
fraction $\delta=0.85 \times 10^{-6}$ the mass of the planet) and about the impactor size cut-offs
play a more significant role than in the calculation of $f_{\rm v}$, as described below.

To explain the results in the middle panels of Fig.~\ref{fig:outcome1}, and to scale these to situations with
different assumptions, note that the two terms on the right hand side of eq.~\ref{eq:dmmdmac} come from
eq.~\ref{eq:matmlosstot} and the top panels of Fig.~\ref{fig:outcome1}, respectively.
The second term explains the most prominent feature on the middle panels of Fig.~\ref{fig:outcome1} which,
as noted already, is that whether an atmosphere grows or shrinks with time is dictated by the $f_{\rm v}$ factor.
That is, the region where planetary atmospheres grow in collisions (dashed lines, darker shading)
is separated from that where they deplete (dotted lines, lighter shading) by the solid $f_{\rm v}=1$ line
(the impact shoreline),
the location of which has all of the dependencies discussed in \S \ref{sss:fvchanges}.

Equations~\ref{eq:eta}, \ref{eq:a} and \ref{eq:vimpvesc} show that
\begin{eqnarray}
  m_{\rm atmloss}/m_{\rm ac} & \propto &
     [D_{\rm max}^{4-\alpha}-D_{\rm min}^{4-\alpha}]^{-1}
     M_\star^\frac{\alpha-1}{3} L_\star^\frac{4-\alpha}{6} M_{\rm p}^\frac{-\alpha-2}{9}
    a_{\rm p}^{-1}
    \times \nonumber \\ & &
    \rho_{\rm p}^\frac{4\alpha-19}{9}
    \delta^\frac{4-\alpha}{3}
    \mu^\frac{2\alpha-8}{3}
    \xi^\frac{2\alpha-2}{3}
    (1+\rho_{\rm p}/\rho_{\rm imp})^\frac{4-\alpha}{3}.
  \label{eq:matmlossmac}
\end{eqnarray}
Since for atmospheres that deplete in collisions $\Delta m/\Delta m_{\rm ac} \approx -m_{\rm atmloss}/m_{\rm ac}$, this
means that the contours in the lighter shaded region would be expected to lie along lines of $M_{\rm p} \propto a_{\rm p}^\frac{-9}{\alpha+2}$,
which for the size distribution assumed in Fig.~\ref{fig:outcome1} are only slightly steeper than the $f_{\rm v}=1$ line.
For planets that are far enough to the left of the $f_{\rm v}=1$ line (i.e., small close-in planets), their large impact velocity
means that impactors are able to remove more atmosphere mass than the planetesimal mass that is accreted.
However, for the known exoplanets the decrease in atmosphere mass is less than the mass that is accreted.

For atmospheres that grow in collisions, $\Delta m/\Delta m_{\rm ac} \approx m_{\rm impacc}/m_{\rm ac}$, which
has a similar scaling to eq.~\ref{eq:matmlossmac} but with some slightly different exponents so that this is
$\propto M_\star^\frac{\alpha-4}{3} M_{\rm p}^\frac{4-\alpha}{9} a_{\rm p}^{0} \rho_{\rm p}^\frac{4\alpha-16}{9}$.
This explains why the contours of constant $\Delta m/\Delta m_{\rm ac}$ become flatter in the darker shaded region,
and moreover there is little dependence on planet mass. 
Indeed, the atmosphere mass gain per impactor mass accreted reaches a plateau in the upper right of the middle panels
of Fig.~\ref{fig:outcome1} at a value which is below $p_{\rm v}$ (which is the maximum possible since this would require all of the volatiles
accreted to go into the atmosphere) by a factor that accounts for the fraction of the impactor mass that arrives
in planetesimals that are too large to be retained in the atmosphere.

To rescale the middle panels of Fig.~\ref{fig:outcome1} for different assumptions, first note that some of the parameters in the model do not
affect the factor $f_{\rm v}$ and so their effect on eq.~\ref{eq:dmmdmac} is relatively straight-forward to determine.
For example, as long as the upper size cut-off does not affect the calculation of $f_{\rm v}$ 
(i.e., as long as the limits in the size distribution do not contribute
to the integrals in equations~\ref{eq:matmlosstot} and \ref{eq:mimpacctot}, see discussion in \S \ref{ss:mult})
then $\Delta m/\Delta m_{\rm ac}$ scales with $\delta$, $\mu$ and $D_{\rm max}$ in the same way
as eq.~\ref{eq:matmlossmac}, i.e.,
\begin{equation}
  \Delta m/\Delta m_{\rm ac} \propto \delta^\frac{4-\alpha}{3} \mu^\frac{2\alpha-8}{3} D_{\rm max}^{\alpha-4},
  \label{eq:dmmdmdep}
\end{equation}
where the dependence on $D_{\rm max}$ has assumed that $\alpha <4$.
This means that atmospheres that are higher in mass have correspondingly larger changes (or need to accrete more
for the same fractional change), as do those that have a more primordial composition
(by a factor of 2.3 when changing from the $\mu_\oplus$ assumed in Fig.~\ref{fig:outcome1} to $\mu_\odot$).
Changing $D_{\rm max}$ can also have a significant effect,
because this affects the fraction of the mass that is in the damaging km-sized planetesimal range,
noting however that there may be an additional $D_{\rm max}$ dependence not accounted for in eq.~\ref{eq:dmmdmdep} if
this affects the integral in eq.~\ref{eq:matmlosstot}.
While there are significant differences for comparable planets between different impactor types, many of these
differences can be understood from the location of the $f_{\rm v}=1$ impact shoreline on the top panels
of Fig.~\ref{fig:outcome1} (see also Fig.~\ref{fig:fvch}).

The bottom panels of Fig.~\ref{fig:outcome1} show the same information as in the middle panels, but this time recording
the fractional change in the planet's atmosphere that would result from accretion of
$\Delta m_{\rm ac} = \Delta m_{\rm ac,LHB}=3 \times 10^{-5}{\rm M}_\oplus$ \citep[i.e., similar to the mass accreted by both the
Earth and Mars during the Late Heavy Bombardment;][]{Gomes2005};
i.e., these panels show $(\Delta m/m)(\Delta m_{\rm ac,LHB}/\Delta m_{\rm ac})$.
This is intended to give the reader an idea of whether impacts are likely to have a significant effect on a planet's atmosphere
following an epoch of heavy bombardment (although as we will describe below, planets in other systems may experience
levels of bombardment that are significantly greater than this, in which case the values in this plot could be scaled accordingly).
This shows that for planets with atmospheres that are expected to deplete in impacts (in the lighter shaded region), it is
relatively easy to deplete these significantly (i.e., to result in $-\Delta m/m$ of order unity or greater). 
For planets that are expected to grow in impacts (in the darker shaded region), growth can be more modest unless the
bombardment was greater than that experienced by the Earth during the Late Heavy Bombardment.

\subsection{Cumulative accreted impactor mass per cumulative incoming mass}
\label{ss:accvsinc}
For a given impactor population (i.e., the incoming planetesimals that have been placed on planet-crossing
orbits with a mass $m_{\rm inc}$), it might be expected that planets in different
regions of the parameter space on Fig.~\ref{fig:outcome1} would end up accreting
different masses (i.e., have a different $m_{\rm ac}$).
Thus a planet that may appear susceptible to atmosphere growth because of a large positive $f_{\rm v}$ in the top panels
Fig.~\ref{fig:outcome1}, and a correspondingly large positive $\Delta m/\Delta m_{\rm ac}$ on the middle panels of
Fig.~\ref{fig:outcome1}, may not grow significantly because it has a low efficiency of accreting the
planetesimals that were placed on planet-crossing orbits.

There are two main considerations here.
First is that planetesimals encountering planets for which $v_{\rm esc} \gg v_{\rm p}$ are more
likely to be ejected in that encounter than to collide with the planet \citep[e.g.,][]{Wyatt2017}. 
Similarly, the timescale for planetesimals to collide with planets that are low in mass (or far from the star)
can be longer than their dynamical lifetime $t_{\rm dyn}$, i.e., the time before which other perturbations
remove the planetesimals from planet-crossing orbits (which may be the same perturbations that put them on planet-crossing
orbits in the first place, like those from more distant planets or stellar companions).
Both effects would result in a low collision efficiency (i.e., a low $m_{\rm ac}/m_{\rm inc}$), and
are hard to quantify because this requires consideration of the other planets in the system that
is better suited to study using N-body simulations than analytics \citep[e.g.,][]{Marino2018, Kral2018}. 

We could make some progress by deriving a rate at which the planetesimals collide with the planet
$R_{\rm ac}$, the rate at which the planet ejects the planetesimals $R_{\rm ej}$ and assuming some fixed 
dynamical loss rate $R_{\rm dyn}$ (that is set by the other perturbers in the system).
The fraction of the impactor population
that is accreted would then be $m_{\rm ac}/m_{\rm inc} = R_{\rm ac}/(R_{\rm ac}+R_{\rm ej}+R_{\rm dyn})$.
Indeed it is possible to derive $R_{\rm ac}$ and $R_{\rm ej}$ for assumptions about the planetesimal orbit
\citep[see][]{Kral2018}. 
However, we refrain from repeating such calculations, since they still require further assumptions
about the specific scenario which would obfuscate the generality of what we are trying to achieve here.
Instead, we plot a few lines on Fig.~\ref{fig:outcome1} which show
for which planets efficiency might be expected to be low.
One of these is $v_{\rm esc}=v_{\rm p}$ (the dark green line on Fig.~\ref{fig:outcome1}),
above which ejection starts to dominate over accretion, which is given by
\begin{equation}
  M_{\rm p} = 40M_\star^{3/2}a_{\rm p}^{-3/2}\rho_{\rm p}^{-1/2}.
  \label{eq:mpejdom}
\end{equation}
The others (the light green lines on Fig.~\ref{fig:outcome1}) are lines of constant accretion time
$t_{\rm acc}$, calculated assuming that planetesimals interact near the pericentres of their
high eccentricity and low inclination ($\sim 0.1$ rad) orbits with a planet on a circular orbit,
which are given by
\begin{equation}
  M_{\rm p} = 30M_\star^{-3/4}a_{\rm p}^3\rho_{\rm p}Q^{9/4}t_{\rm acc}^{-3/2},
  \label{eq:mpdyndom}
\end{equation}
where $t_{\rm acc}$ is in Myr and $Q$ is the planetesimals' apocentre distance in au which is assumed to
be $10a_{\rm p}$ in the figures.
Dynamical removal starts to dominate over accretion below the line for which $t_{\rm acc}=t_{\rm dyn}$
(or equivalently, accretion efficiency drops by a factor $\sim t_{\rm dyn}/t_{\rm acc}$).

The lines of eqs.~\ref{eq:mpejdom} and \ref{eq:mpdyndom} on Fig.~\ref{fig:outcome1} are only meant as a guide,
and do not delineate those planets that do and those that do not suffer impacts.
For example, while the accretion time for the Earth is $\sim 1$\,Gyr and so $3-4$ orders of magnitude longer than the
typical dynamical lifetime of comets in the inner Solar system of $\sim 0.3$\,Myr \citep[][]{Levison1997},
it was still able to accrete $3\times 10^{-5}$\,${\rm M}_\oplus$ during the Late Heavy Bombardment \citep{Gomes2005}.
This is because the low accretion efficiency $\sim 10^{-6}$ was overcome by a large mass of planetesimals undergoing
scattering during this event \citep[$\sim 30$\,${\rm M}_\oplus$,][]{Gomes2005}.
Systems with more regularly spaced planets have higher accretion efficiencies \citep[$\sim 1$\%, e.g.,][]{Marino2018},
and so can undergo significant accretion without requiring such a major upheaval as the Late Heavy Bombardment.
That is, these lines cannot account for the fact that the mass accreted also 
depends on the ability of external planets to put planetesimals on such orbits among other factors.
Nevertheless these lines show that small planets that are close to the star should have a
high collision efficiency, since they might be expected to accrete most planetesimals that are put on planet-crossing orbits,
with the caveat that accretion efficiency might still be low if a planet is competing with other nearby planets
that also have high accretion efficiencies \citep[as in the TRAPPIST-1 system,][]{Kral2018}.

\subsection{Dependence on stellar mass}
\label{ss:mstar}

\begin{figure*}
  \begin{center}
    \vspace{-0.05in}
    \begin{tabular}{cc}
      \hspace{-0.1in} Asteroidal Impactors & \hspace{-0.5in} Cometary Impactors \\[-0.6in]
      \hspace{-0.45in} \includegraphics[width=1.25\columnwidth]{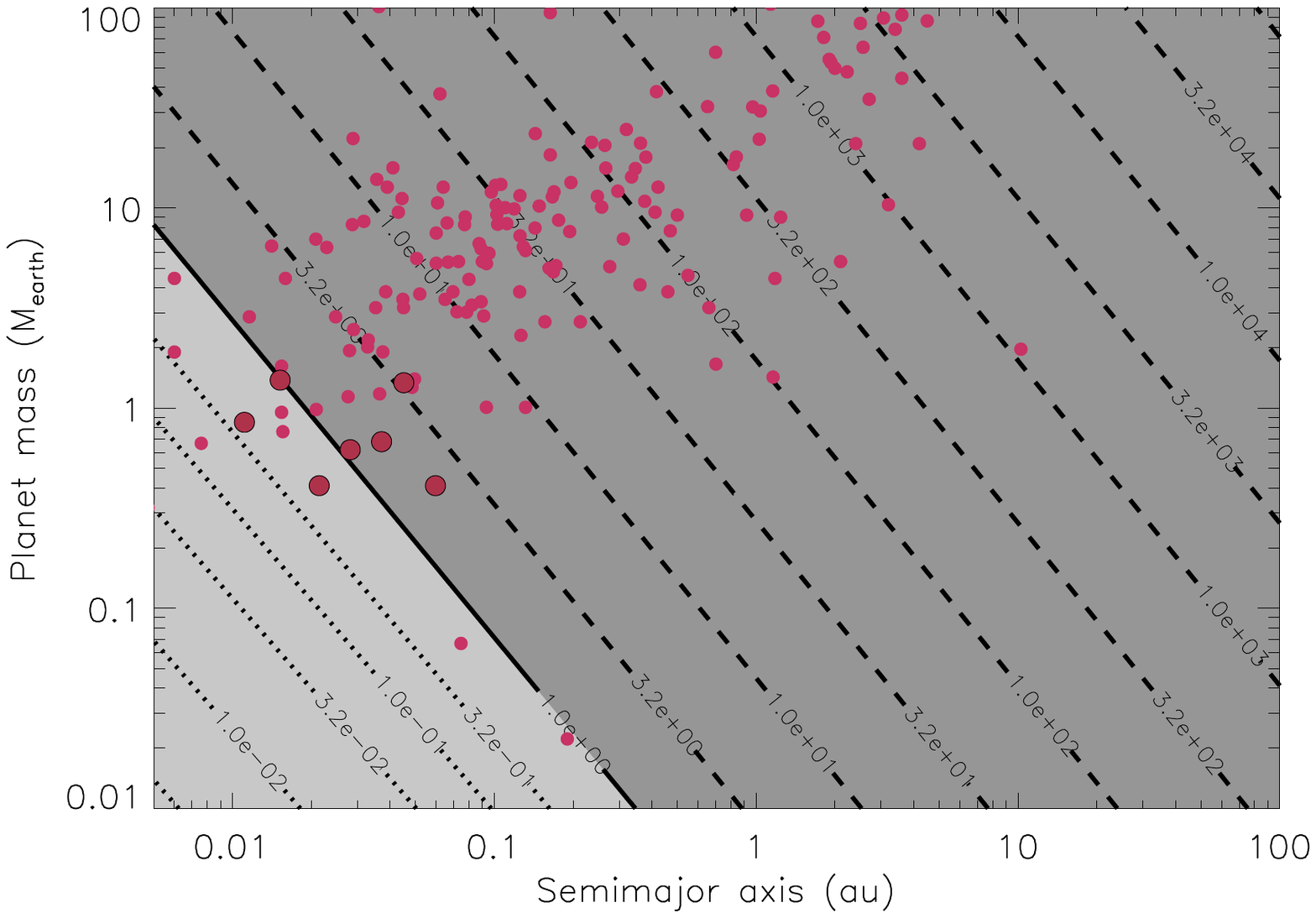} &
      \hspace{-0.85in} \includegraphics[width=1.25\columnwidth]{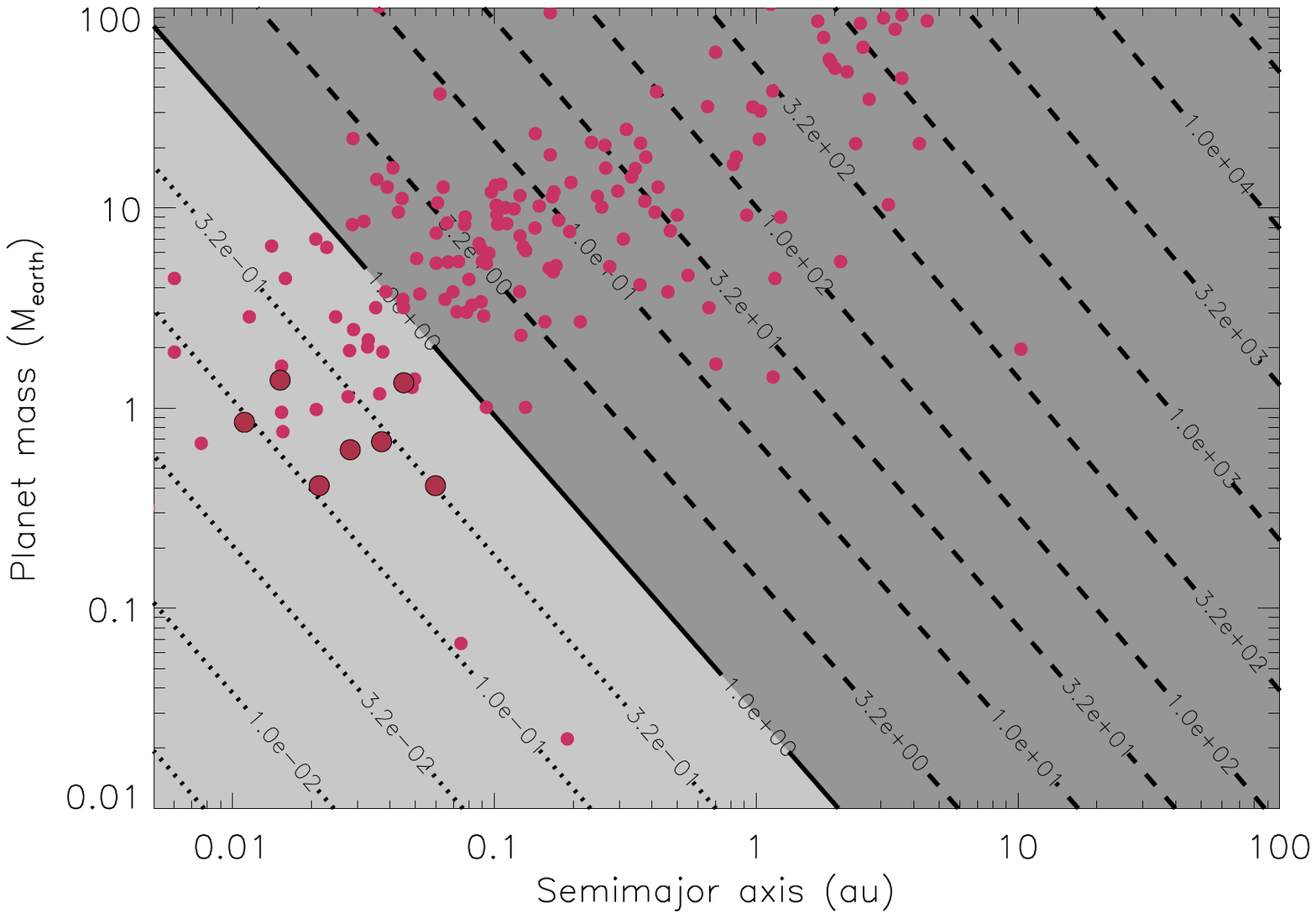} \\[-3.05in]
      \hspace{-0.45in} \includegraphics[width=1.25\columnwidth]{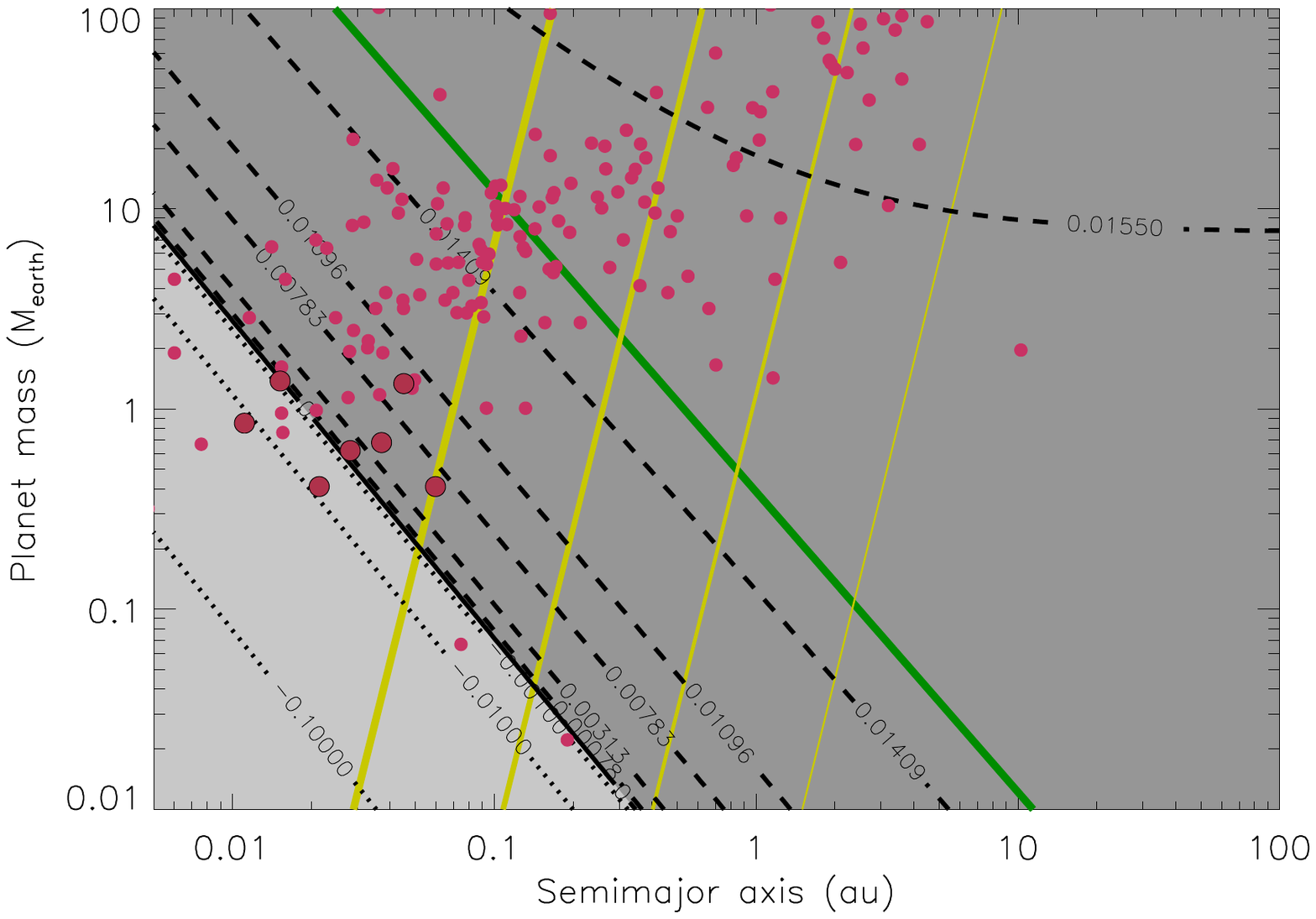} &
      \hspace{-0.85in} \includegraphics[width=1.25\columnwidth]{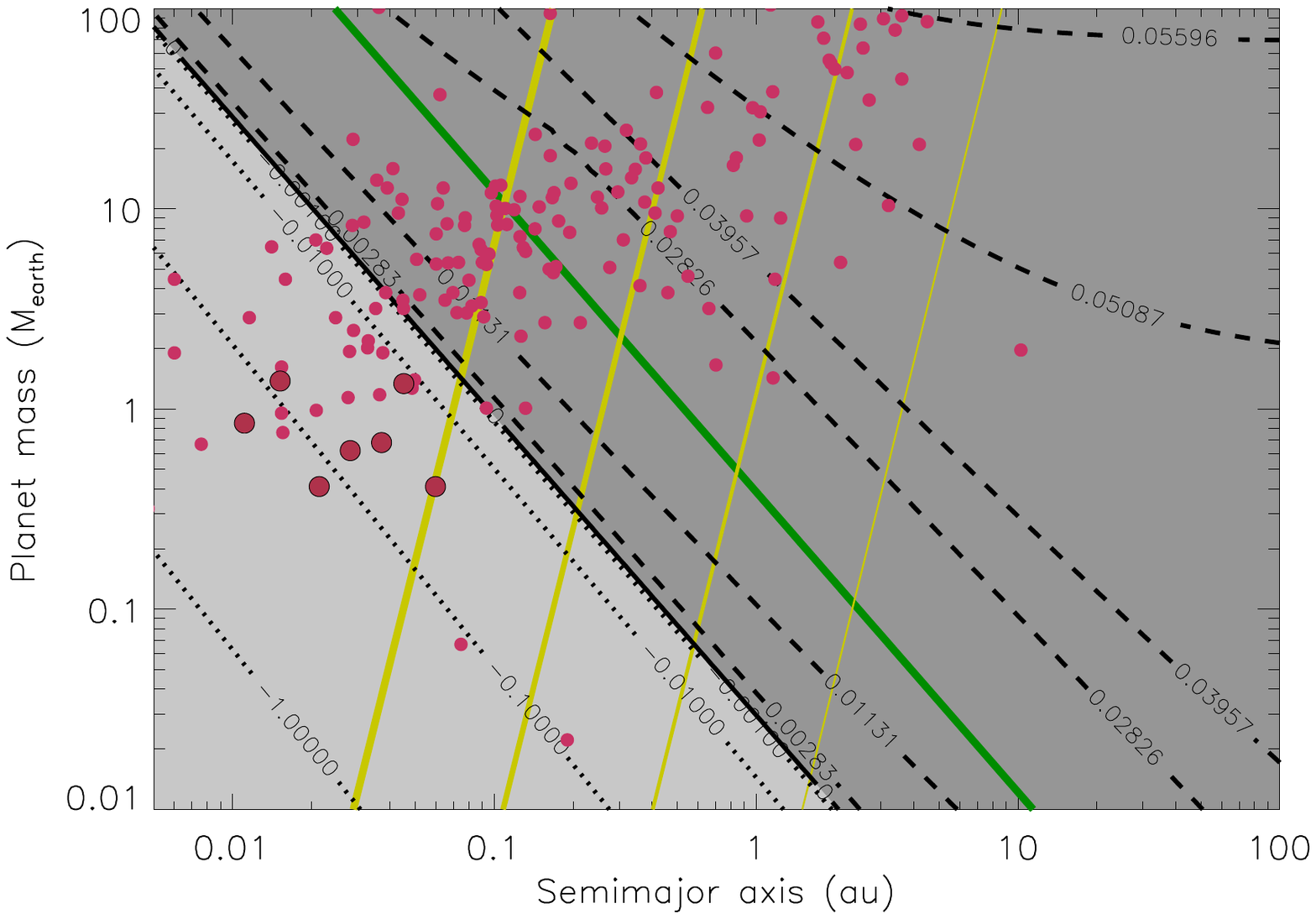} \\[-3.05in] 
      \hspace{-0.45in} \includegraphics[width=1.25\columnwidth]{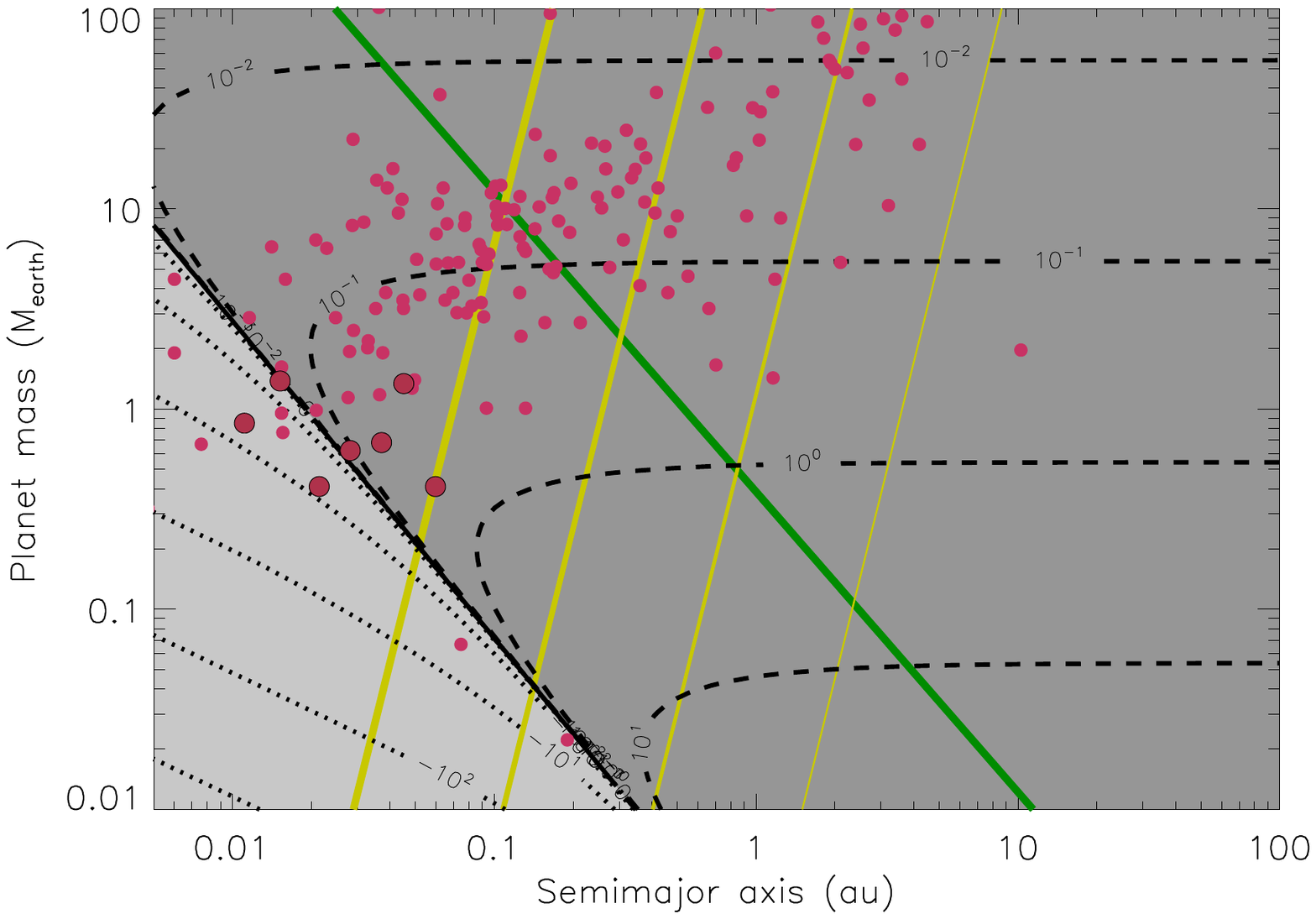} &
      \hspace{-0.85in} \includegraphics[width=1.25\columnwidth]{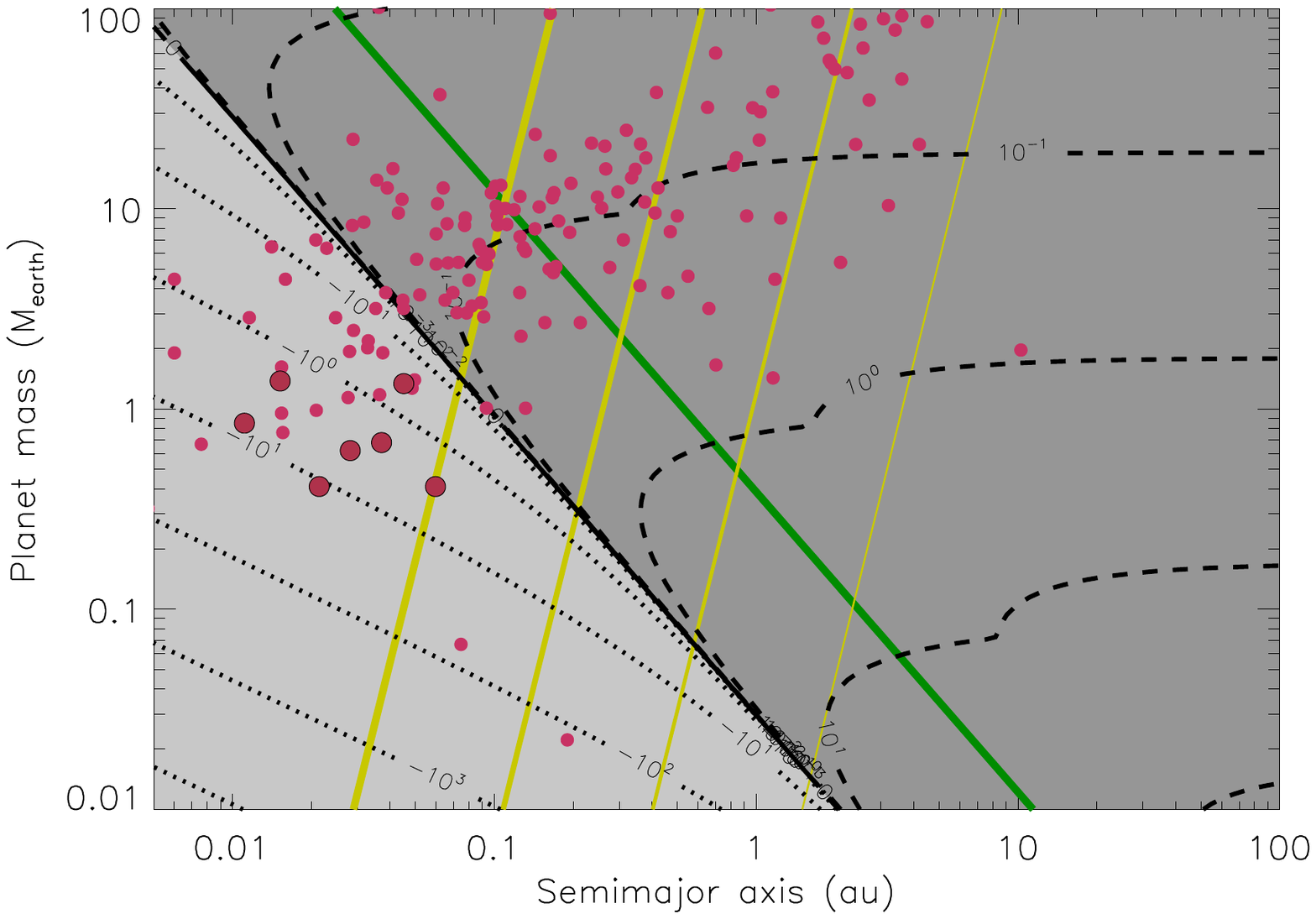} 
    \end{tabular}
    \vspace{-2.55in}
    \caption{As for Fig.~\ref{fig:outcome1} but for planets orbiting $0.08{\rm M}_\odot$ stars.
    Here the red circles are known exoplanets for $<0.6{\rm M}_\odot$ stars
    \citep[from the exoplanet.eu database on 28 November 2018,][]{Schneider2011}, with the
    7 planets in the TRAPPIST-1 system highlighted by the larger symbols \citep[with parameters from][]{Gillon2017}.
    }
   \label{fig:outcome2}
  \end{center}
\end{figure*}

Fig.~\ref{fig:outcome2} shows the same calculations as for Fig.~\ref{fig:outcome1}, but this
time for planets orbiting stars with $M_\star=0.08{\rm M}_\odot$ and $L_\star=5.2 \times 10^{-4}{\rm L}_\odot$,
i.e., with parameters appropriate for the TRAPPIST-1 system \citep{Gillon2017}. 
Comparison of the top panels in the two figures shows how the slower orbital velocity (and so smaller impact velocity) for
lower mass stars results in less destructive impacts for planets with the same properties.
Nevertheless, the location of the $f_{\rm v}=1$ line explains why \citet{Kral2018} concluded that the
closest in planets in the TRAPPIST-1 system would have their atmospheres stripped in cometary impacts.
Their conclusion that the atmospheres of the outermost planets would grow in collisions is because their calculations
made different assumptions about the distribution of impact velocities (which are more realistic for the scenario they
were considering for this system).

\section{Discussion}
\label{s:disc}
This paper has considered the effect of planetesimal impacts on planetary atmospheres, using assumptions that are
valid when the atmosphere is not massive enough for planetesimals to disintegrate before reaching the surface,
or for the structure of the atmosphere to deviate from our simple prescription, and (justifiably) ignoring the effect of giant impacts.
Some starting point for the atmosphere has been assumed, and other factors which may affect the evolution of the
atmosphere are ignored, such as photoevaporation due to stellar photons or outgassing of volatiles from the interior that
were inherited during formation.
While these caveats should be born in mind in the following, these assumptions make it possible to draw some
broad conclusions about the effect of planetesimal impacts on planetary atmospheres that are summarised in \S \ref{ss:summary}
before considering how giant impacts or massive atmospheres might affect those conclusions in \S \ref{ss:stochasticity}-\ref{ss:massive},
then going on to consider the implications for specific systems in \S \ref{ss:ss}-\ref{ss:pop}, as well as the broader implications for
the development of life in \S \ref{ss:life}.

\subsection{Summary}
\label{ss:summary}
\begin{figure}
  \begin{center}
    \vspace{-0.55in}
    \begin{tabular}{c}
      \hspace{-0.5in} \includegraphics[width=1.25\columnwidth]{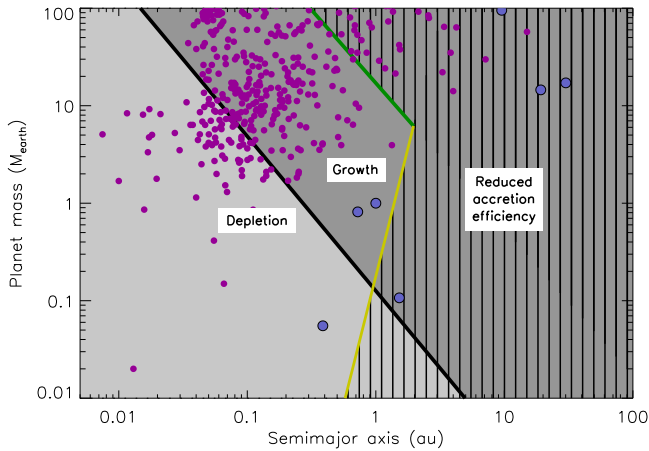}
    \end{tabular}
    \vspace{-2.55in}
    \caption{Summary of the different outcomes of bombardment that might be expected for the atmospheres of planets
    in different regions of parameter space, for planets orbiting solar mass stars being impacted by asteroidal impactors.
    The two main regions are that of atmosphere growth (darker shaded region) and depletion (lighter shaded region) that are
    divided by the impact shoreline shown with the thick black line.
    However, in the cross-hatched region a reduced accretion efficiency might lead to less change in atmosphere.
    }
   \label{fig:schematic}
  \end{center}
\end{figure}

The main conclusion of \S \ref{s:comp} is that the planet mass - semimajor axis parameter
space can be divided into regions with different outcomes,
with some dependence on stellar mass and on the physical and dynamical properties of the impactors.
This is illustrated in Fig.~\ref{fig:schematic} which shows lines appropriate for asteroidal impactors onto
planets orbiting solar mass stars, noting that the boundaries between the different regions are not meant to be strictly
interpreted.

\subsubsection{Planets expected to have no atmosphere (region labelled depletion)}
\label{sss:noatm}
Planets that have $f_{\rm v}<1$ and $t_{\rm acc}\ll 3$\,Gyr would be
expected to have any primordial atmosphere depleted by bombardment.
This applies to planets that are both low in mass and very close to their host stars, a prime example being
the innermost planets orbiting TRAPPIST-1 \citep{Kral2018}.
The low negative values of $\Delta m/m$ following accretion of $3 \times 10^{-5}{\rm M}_\oplus$ in this regime shown on the bottom panels of
Figs.~\ref{fig:outcome1} and \ref{fig:outcome2} mean that these planets could be expected to completely lose any Earth-like
atmospheres when subjected to bombardment levels comparable to that inferred for the Earth during the Late Heavy Bombardment.
The bombardment level required for complete atmosphere loss
can be inferred from the middle panels of Figs.~\ref{fig:outcome1} and \ref{fig:outcome2}, since eq.~\ref{eq:macbare} shows that
\begin{equation}
  \Delta m_{\rm ac,bare}/m_0 = 3(\alpha-1)^{-1}(\Delta m/\Delta m_{\rm ac})^{-1},
  \label{eq:dmacbare}
\end{equation}
i.e., the mass that needs to be accreted is approximately the atmosphere mass divided by the value
plotted in those panels (noting that eq.~\ref{eq:dmmdmdep} shows that the
plotted value would also need to be scaled by $[\delta_0/0.85 \times 10^{-6}]^{0.17}$).
The only impediment to these planets having completely lost their atmospheres is either an absence of
impactors (i.e., below a level given by the initial atmosphere mass divided by the value plotted
in the middle panel of Fig.~\ref{fig:outcome1}), or for the initial atmospheres to be sufficiently massive
(although in such extremes the assumptions in this paper might break down, see \S \ref{ss:massive}).

\subsubsection{Planets expected to have atmospheres enhanced in collisions (region labelled growth)}
\label{sss:atmgrowth}
Planets for which $f_{\rm v}>1$ and $t_{\rm acc} \ll 3$\,Gyr and $v_{\rm esc} < v_{\rm p}$
would be expected to grow secondary atmospheres in collisions.
This applies to planets that are close to the star, more massive than those depleted in collisions discussed in \S \ref{sss:noatm},
but not so massive that their large escape velocity results in a reduced accretion efficiency.
There still needs to be a sufficient level of bombardment for the atmospheres to grow significantly, but the bottom panels of
Figs.~\ref{fig:outcome1} and \ref{fig:outcome2} show that
slightly higher than Late Heavy Bombardment-levels of accretion would be sufficient to grow an
Earth-like atmosphere (in the sense that $\delta=\delta_\oplus$) for many such planets.
The middle panels of Figs.~\ref{fig:outcome1} and \ref{fig:outcome2}
suggest that atmospheres could grow in mass by typically $\sim 1$\% of the impactor mass accreted.
Thus the 1\% accretion efficiency seen in the simulations of \citet{Marino2018} could result in atmospheres 100 times more massive
than that on Earth for bombardment involving just $1{\rm M}_\oplus$ of planetesimals, which could be a fraction of any
planetesimal belt.

\subsubsection{Planets likely unaffected by collisions (region labelled reduced accretion efficiency)}
\label{sss:unaffected}
The atmospheres of planets that are either far from the star, or very high in mass, may be largely unaffected by collisions.
This is not because they would be unaffected by any collisions that occurred. 
Indeed atmosphere growth or depletion is always the favoured outcome in the darker and lighter shaded regions of Fig.~\ref{fig:schematic}
(with the caveat that this boundary has some uncertainties as noted in Fig.~\ref{fig:fvch}).
Rather this is because planetesimals could be removed dynamically from the planet's vicinity faster than they can undergo collisions,
resulting in a low accretion efficiency.
Planets that are susceptible to having a low accretion efficiency are identified by having
$v_{\rm esc}>v_{\rm p}$ and/or $t_{\rm acc} \gg 3$\,Gyr.
However, it is important to emphasise the caveat that such dynamical removal depends on what other planets are present in the
system, and it could be that planets in this region still manage to accrete a significant quantity of planetesimals and so
have their atmospheres altered in the way indicated by the shading.

\subsection{Giant impacts}
\label{ss:stochasticity}
The parameterisation for $\chi_{\rm a}$ in eq.~\ref{eq:matmloss} is not applicable to giant impacts for
which a planet's atmosphere is not only lost locally at the point where the impact occurs.
Rather giant impacts send a shock wave through the body of the planet, which is transmitted to the atmosphere.
This can accelerate parts of the atmosphere to beyond the escape velocity, leading to partial loss of the atmosphere globally.
A prescription for the outcome of giant impacts is that the atmospheric mass lost per impactor mass 
can be approximated for an isothermal atmosphere by \citep{Schlichting2015}
\begin{equation}
  m_\mathrm{atmloss,GI}( x ) / m_{\rm imp} = \delta ( v_{\rm imp}/v_{\rm esc} ) [ 0.4 + 1.4 x - 0.8 x^2 ],
  \label{eq:gi_x_loss}
\end{equation}
where $x \equiv (v_{\rm imp}/v_{\rm esc}) (m_{\rm imp}/M_{\rm p})$;
for an adiabatic atmosphere, the coefficients are instead 0.4, 1.8, and $-$1.2.

This means that the shock wave caused by a giant impact results in an atmospheric mass loss
per unit impactor mass that typically remains constant (i.e., independent of impactor size) up to very large impactors,
at a level that is proportional to the atmosphere to planet mass ratio $\delta$ times the ratio of impact to escape
velocities $(v_{\rm imp}/v_{\rm esc})$.
This should be added to the local atmospheric mass loss plotted in Fig.~\ref{fig:mm} which
in constrast decreases rapidly with increasing impactor size.
This means that there is a size $D_{\rm GI}$ above which giant impacts dominate atmospheric mass loss, and
below which giant impact erosion can effectively be ignored.
This transition can be calculated by equating $m_{\rm atmloss,GI}$ from eq.~\ref{eq:gi_x_loss} with $m_{\rm atmloss}$
from eq.~\ref{eq:matmloss}.
However, to give the reader a feeling for where this transition occurs note that
the prescription from \citet{Schlichting2015} \citep[which is similar but not identical to that of][]{Shuvalov2009}
puts the boundary at approximately
\begin{equation}
  D_\mathrm{GI}
    \simeq
      \left[
        1.6 H R_\mathrm{p}^2 \,
        \left( v_\mathrm{esc} / v_\mathrm{imp} \right)
        \left( \rho_\mathrm{p} / \rho_\mathrm{imp} \right)
      \right]^{1/3}.
\end{equation}

The combined effect of multiple giant impacts can be computed by integrating
$ m_\mathrm{atmloss,GI}( x ) / m_\mathrm{imp} $
over the size distribution of the bodies causing giant impacts (under the assumption that these arrive in steady state).
Using the assumed power-law size distribution, the atmospheric mass loss per unit impactor mass is
\begin{align}
& \frac{ m_\mathrm{atmloss,GI} }{ m_\mathrm{ac} }
  =
    \delta \,
    \frac{ v_\mathrm{imp} }{ v_\mathrm{esc} }
    \Bigg\{
      0.4
\nonumber \\
  & \quad
      + 1.4
        \left( \frac{ 4 - \alpha }{ 7 - \alpha } \right)
        \left[
          \frac{ x_\mathrm{max}^{ ( 7 - \alpha ) / 3 } - x_\mathrm{min}^{ ( 7 - \alpha ) / 3 } }
            { x_\mathrm{max}^{ ( 4 - \alpha ) / 3 } - x_\mathrm{min}^{ ( 4 - \alpha ) / 3 } }
        \right]
\nonumber \\
  & \quad
      - 0.8
        \left( \frac{ 4 - \alpha }{ 10 - \alpha } \right)
        \left[
          \frac{ x_\mathrm{max}^{ ( 10 - \alpha ) / 3 } - x_\mathrm{min}^{ ( 10 - \alpha ) / 3 } }
            { x_\mathrm{max}^{ ( 4 - \alpha ) / 3 } - x_\mathrm{min}^{ ( 4 - \alpha ) / 3 } }
        \right]
    \Bigg\},
    \label{eq:matmlossgi}
\end{align}
which works for all power-law indices except $ \alpha = 4 $, $ \alpha = 7 $, and $ \alpha = 10 $.

\begin{figure}
  \begin{center}
    \vspace{-0.52in}
    \begin{tabular}{c}
      \hspace{-0.5in} \includegraphics[width=1.25\columnwidth]{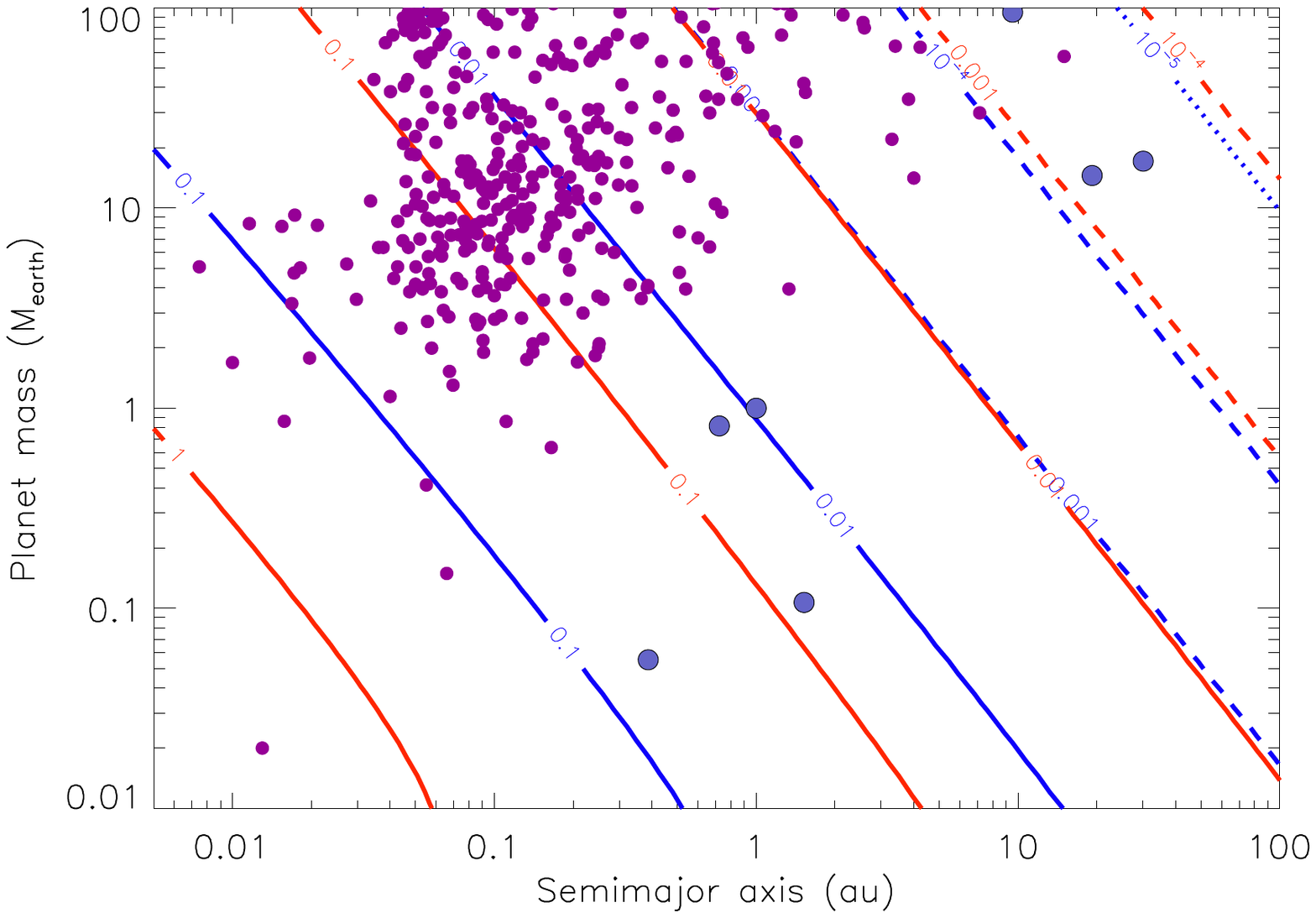} \\[-3.05in]
      \hspace{-0.5in} \includegraphics[width=1.25\columnwidth]{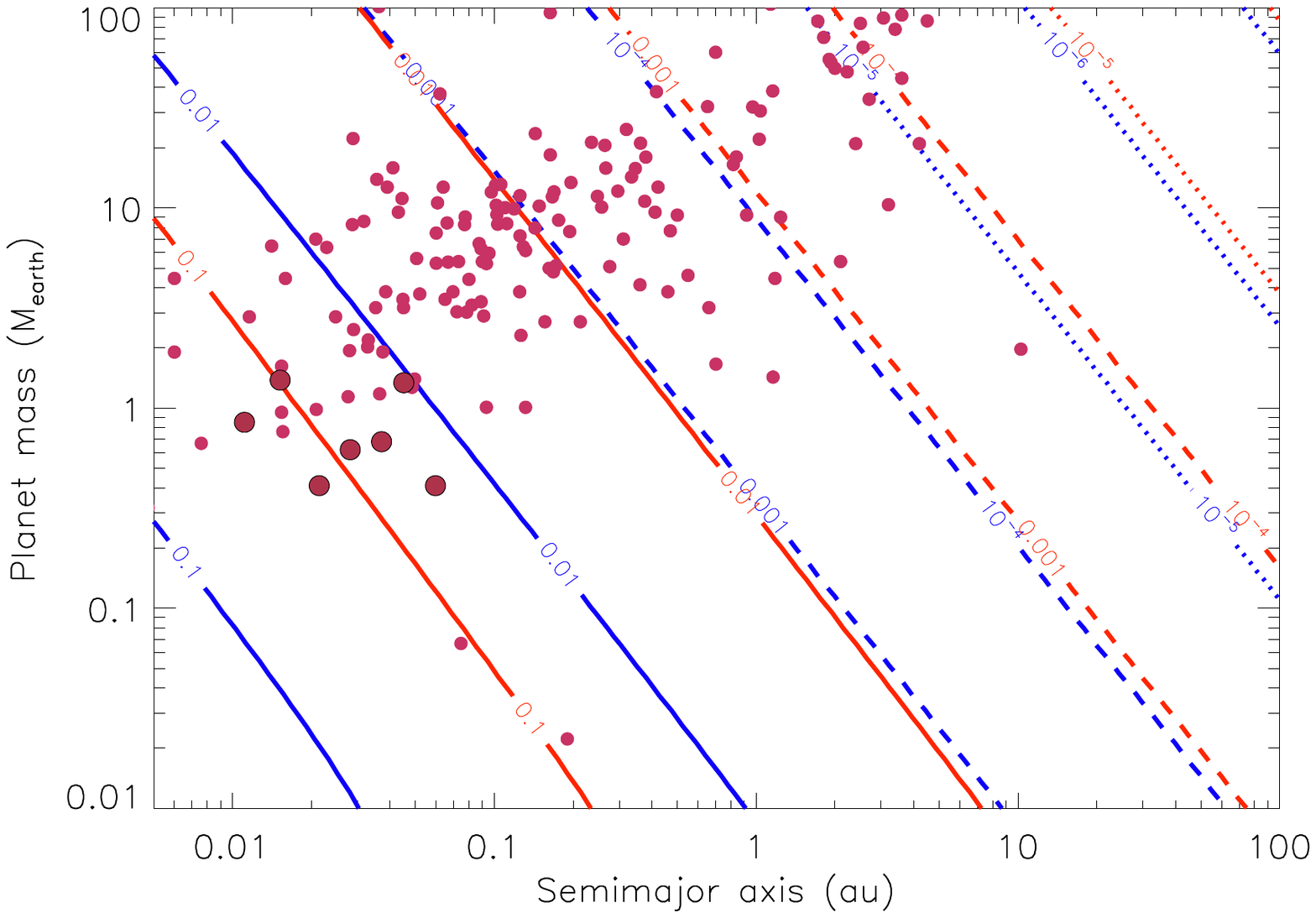}
    \end{tabular}
    \vspace{-2.5in}
    \caption{
    Atmosphere-to-planet mass ratio $\delta_{\rm GI}$ above which giant impacts
    dominate an atmosphere's mass loss over the local effects of smaller impacts
    (i.e., the line shows where $m_{\rm atmloss,GI}(\delta_{\rm GI}) = m_{\rm atmloss}(\delta_{\rm GI})$).
    Both panels assume an impactor size distribution with $\alpha = 3.5$
    that extends from $D_{\rm min}=1$\,m to $D_{\rm max} = 1000$\,km,
    a planet density of $\rho_{\rm p} = 5.5$\,g\,cm$^{-3}$,
    and an atmosphere with mean molecular weight $\mu = 29$.
    Asteroidal impactors are shown in blue ($\rho_{\rm imp} = 2.8$\,g\,cm$^{-3}$ and $\xi=0.3$),
    and cometary impactors in red ($\rho_{\rm imp} = 0.9$\,g\,cm$^{-3}$ and $\xi=1$).
    The top panel assumes a star with $M_\star = 1\,{\rm M}_\odot$ and $L_\star = 1\,{\rm L}_\odot$,
    while the bottom panel uses $M_\star = 0.08\,{\rm M}_\odot$ and $L_\star = 5.2 \times 10^{-4}\,{\rm L}_\odot$.
    The line-style is simply a function of $\delta_{\rm GI}$, with the solid and dashed lines indicating that
    giant impacts only dominate in atmospheres that are sufficiently massive for the assumptions in the
    model to break down.
    }
   \label{fig:deltagi}
  \end{center}
\end{figure}

To quantify the regime where it is no longer possible to ignore giant impact induced atmospheric mass loss, 
Figure~\ref{fig:deltagi} shows the atmosphere to planet mass ratio $\delta_{\rm GI}$ at which giant impact
mass loss (eq.~\ref{eq:matmlossgi}) is equal to that caused by local effects (eq.~\ref{eq:matmlosstot}).
Unlike Fig.~\ref{fig:mm} for which the $A$ factors from eq.~\ref{eq:matmlosstot} and \ref{eq:mimpacctot} cancelled,
Fig.~\ref{fig:deltagi} has had to make assumptions about the star, planet and impactors which are noted in the caption.
Nevertheless, these plots show that atmospheres have to be a substantial fraction of
the planet's mass before giant impact induced atmospheric mass loss becomes important,
with high mass planets at large distances from the star being most susceptible to such effects, primarily because
of the small relative velocity of impacts in this region.
Note that planets in this regime were expected to grow by impacts when giant impacts were ignored
(see top panels of Figs.~\ref{fig:outcome1} and \ref{fig:outcome2}), so while including giant
impacts into the analysis would have the effect of reducing $f_{\rm v}$, this would not necessarily
reverse the conclusion that impacts would result in the atmospheric growth for such planets.
Asteroidal (rather than cometary) impactors also have a greater propensity for atmospheric loss by giant
impacts, as do planets around lower mass stars.
A planet like the Earth would require its atmosphere to be of order 1\% of the planet mass before
giant impacts become important.
This explains why \citet{Schlichting2015} concluded that giant impacts do not dominate atmosphere erosion, which
holds as long as the atmosphere is not too massive.

Individual impacts can have a devastating effect on an atmosphere.
This becomes the case when the mass lost in an individual impact is of order the atmosphere mass, which occurs when $x \approx 1$
for the prescription of eq.~\ref{eq:gi_x_loss} (above which the prescription is no longer valid).
Thus invidual impactors can only be ignored when the largest impactor has a mass
$m_{\rm imp}$ that is much less than $M_{\rm p}v_{\rm esc}/v_{\rm imp}$.
That is, the stochastic effect of individual impactors cannot be ignored when
impactor masses are close to the mass of the planet (or indeed much smaller if the impact velocity is large enough),
and this is independent of how massive the atmosphere is. 
While such events may be expected to be inevitably rare for most size distributions, their stochastic nature
could result in an atmospheric mass different from that predicted in Figs.~\ref{fig:outcome1} and \ref{fig:outcome2}, and
in particular this could explain differences in the atmospheres of neighbouring planets which should
have undergone similar bombardment histories, or at least ones that should be different in a predictable way so
that any differences in their atmospheres that result from impacts should be relatively well known
\citep[e.g.,][]{Griffith1995, Biersteker2018}.

To summarise, the effect of giant impacts can be implemented into models of atmospheric evolution using
eq.~\ref{eq:matmlossgi}
\citep[though it may also be important to consider the contribution to the atmosphere from
material vaporised from the planet surface, e.g.,][]{Okeefe1989, Melosh1989b, Vickery1990, Pope1997},
with the further assumption that impactor retention is unaffected by the additional physics of giant impacts
(i.e., this is still given by eq.~\ref{eq:mimpacc}).
The stochastic effect of individual impacts could also be readily included using Monte Carlo methods
\citep[e.g.,][]{Griffith1995, deNiem2012, Wyatt2014}.
However, we conclude that this is unlikely to have a significant effect, except in the case that the atmosphere is
already massive (as quantified in Fig.~\ref{fig:deltagi}), or if the largest impactors are comparable in mass to the planet.
It is, however, worth noting that other authors have inferred giant impacts to play an important role in atmosphere evolution
\citep[e.g.,][]{deNiem2012}.
The explanation for this discrepancy seems to be that those studies extrapolated parameterised outcomes derived for
$<10$\,km bodies \citep{Svetsov2007} up to $>100$\,km bodies for which the relevant physics is different thus
requiring different parameterisation \citep[see][]{Schlichting2015}.
Nevertheless, this highlights that there remain some differences in the literature on the correct approach to modelling the outcomes,
which can result in qualitatively different evolution.

\subsection{Massive atmospheres}
\label{ss:massive}
The prescription for the outcome of impacts used in this paper is valid for impactors that reach the planet's surface.
This is inevitably not the case for the smallest impactors, which instead cause aerial bursts or fragment before reaching
the surface, changing their effect on the atmosphere.
This is particularly relevant for massive atmospheres, like that of Venus, for which this can be relevant for the 10s of km
size range of planetesimals that had been predicted to have most effect on the planet's atmosphere.
Simulations in this regime were performed in \citet{Shuvalov2014}, which also provided a prescription to implement
this in a manner similar to that presented in \S \ref{ss:ind} (see their eqs~7-11).
However, since these simulations were only performed for an Earth-like planet, their equations 9 and 10
were not generalised to the range of planet masses being considered here. 
Nevertheless, their results can be used to give a qualitative understanding of how this would change the results.

The main consequence of aerial bursts is to change Fig.~\ref{fig:mm} in the regime of impactors smaller than a certain
size, which means for $\eta < \eta_{\rm ab}$, where
\begin{equation}
  \eta_{\rm ab}=0.19(\rho_0/\rho_{\rm imp})^{1/2}(1+\rho_{\rm imp}/\rho_{\rm p})^{-1}[(v_{\rm imp}/v_{\rm esc})^2-1].
  \label{eq:etaab}
\end{equation}
There is also a narrow range of $\eta$ for which fragmentation before impact is important, extending from $\eta_{\rm ab}$
up to $\eta_{\rm fr} \approx 4.0\eta_{\rm ab}$.
Since $\eta_{\rm ab}$ has a dependence on the density of the atmosphere, a more massive atmosphere results in larger
planetesimals being affected.
In the regime where aerial bursts are important, this results in an increased atmospheric loss,
i.e., a greater $m_{\rm atmloss}(D)/m_{\rm imp}$, the level of which scales $\propto \eta^{1/3}m^{1/3}$ (among
other dependencies).
That is, the level of mass loss for a given $\eta$ depends on the atmosphere mass, which was not the case before,
adding an additional parameter to be considered in the analysis.
Impactor retention in this regime can be assumed to be 100\%.

It is not the purpose of this paper to explore this in detail, but it is worth noting that this prescription
could mean that atmosphere growth might stall, as atmosphere loss becomes more efficient as the mass grows.

\subsection{Application to the Solar system}
\label{ss:ss}
Our model was already applied in \S \ref{ss:fv} to the question of whether the atmospheres of the terrestrial planets in the Solar system grow or
deplete in planetesimal collisions.
Here we expand on Fig.~\ref{fig:outcome1} to consider the effect of a Late Heavy Bombardment-like bombardment level
on the current atmospheres of the terrestrial planets (i.e., using the actual planet properties rather than
reference values) for the given assumptions about asteroidal or cometary impactors (see Table~\ref{tab:sspl}).
Thus, Earth and Venus atmospheres grow by +39\% and +0.2\% for asteroidal impactors, respectively,
but both deplete in cometary impacts, with Mercury also being depleted in all impacts,
and Mars depleted in cometary impacts but growing its atmosphere for asteroidal impactors.
Further work would be needed to consider the implications of this model for Uranus and Neptune, since while Fig.~\ref{fig:outcome1}
might suggest that neither planet should have their atmospheres significantly enhanced with an LHB-like level of
accretion, that level refers only to that accreted onto the Earth and both planets have $v_{\rm esc} \gg v_{\rm p}$
and long accretion times suggesting a low accretion efficiency, and moreover the ice giants have atmospheres that
are sufficiently massive for the prescription to be invalid.

\begin{table*}
  \centering
  \caption{Properties of Solar system terrestrial planets, and the predictions of the model for the
  fractional change in atmosphere mass due to accretion of $3 \times 10^{-5}{\rm M}_\oplus$ of impactors
  with a size distribution $\alpha=3.5$ from 1\,m up to 100\,km of asteroidal ($\rho_{\rm imp}=2.8$\,g\,cm$^{-3}$,
  $p_{\rm v}=0.02$, $\xi=0.3$) or cometary ($\rho_{\rm imp}=0.9$\,g\,cm$^{-3}$, $\xi=1$) type.}
  \label{tab:sspl}
  \begin{tabular}{cccccccc}
     \hline
     Planet  & $a_{\rm p}$  & $M_{\rm p}$  & $\rho_{\rm p}$  & $\delta$  & $\mu$  &
     $(\Delta m_{\rm LHB,ac}/m)_{\rm ast}$ & $(\Delta m_{\rm LHB,ac}/m)_{\rm com}$ \\
     \hline 
     Venus   & 0.72  & 0.82  & 5.2  & $99 \times 10^{-6}$    & 43.5 & +0.24\% & -1.4\% \\
     Earth   & 1.0   & 1.0   & 5.5  & $0.85 \times 10^{-6}$  & 29.0 & +39\% & -16\% \\
     Mars    & 1.52  & 0.11  & 3.9  & $0.039 \times 10^{-6}$ & 43.3 & -6200\% & -24,000\% \\
     \hline
  \end{tabular}
\end{table*}

\begin{figure}
  \begin{center}
    \vspace{-0.6in}
    \begin{tabular}{c}
      \hspace{-0.55in} \includegraphics[width=1.25\columnwidth]{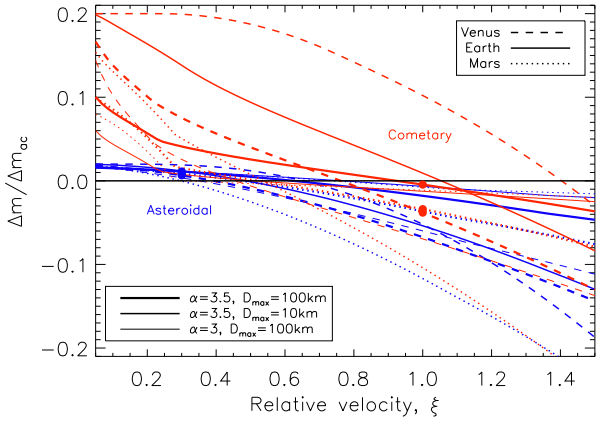}
    \end{tabular}
    \vspace{-2.55in}
    \caption{Change in atmosphere mass per accreted impactor mass for the Solar system terrestrial planets
    assuming their current properties (see Table~\ref{tab:sspl}).
    This is plotted for different assumptions about the impactors with the ratio of the relative velocity of impactors
    to the planet's orbital velocity ($\xi$) on the $x$-axis.
    The size distribution is assumed to be a power law from $D_{\rm min}=1$\,m up to $D_{\rm max}=10$\,km or 100\,km,
    with a slope of $\alpha=3.0$ or 3.5.
    Asteroidal impactors are those with $\rho_{\rm imp}=2.8$\,g\,cm$^{-3}$ and $p_{\rm v}=0.02$ and cometary impactors
    are those with $\rho_{\rm imp}=0.9$\,g\,cm$^{-3}$ and $p_{\rm v}=0.2$.
    The values for the assumptions used elsewhere in the paper are shown with filled circles.
    }
   \label{fig:dmmdmall}
  \end{center}
\end{figure}

However, the discussion in \S \ref{sss:fvchanges} already gives reason for caution when interpreting such values,
since they are highly sensitive to the assumptions.
Here we expand on this point in Fig.~\ref{fig:dmmdmall} which shows how the change in atmosphere mass per impactor mass
accreted (i.e., $\Delta m / \Delta m_{\rm ac}$) depends on assumptions about the impactor relative velocity ($\xi$)
and size distribution ($\alpha$ and $D_{\rm max}$) for asteroidal and cometary impactors (now defined only by
their density and contribution to the atmosphere, $\rho_{\rm imp}$ and $p_{\rm v}$).
This shows how changing the impactor relative velocity from $\xi=0.3$ to 0.5 for asteroidal impactors and from
$\xi=1.0$ to 0.8 for cometary impactors (which as noted in \S \ref{sss:imptypes} may be a more realistic assumption based on
N-body simulations) would have resulted in the opposite conclusion for the Earth, i.e., that the atmosphere
would grow in cometary impacts and deplete in asteroidal impacts.
Similarly the size distribution plays a strong role, with atmosphere growth favoured more for distributions with the
smaller 10\,km upper cut-off.
This is because impacts with 10-100\,km planetesimals destroy atmospheres rather than lead to their growth, so
removing these from the distribution increases $\Delta m/m_{\rm ac}$, although only up to a maximum of $p_{\rm v}$,
which is only reached if all of the accreted planetesimal mass is retained and a negligible fraction of atmosphere lost in
impacts (i.e., for low velocity collisions).
Flattening the size distribution (i.e., the thinner lines with $\alpha=3.0$) has the opposite effect because it then
places more of the mass in larger planetesimals.

Clearly for the Solar system where the size distribution is known for the different impactor populations,
and where these populations also have relative velocities that can be derived from N-body simulations, 
the approach of using a power law size distribution and single $\xi$ value can inevitably only give
an approximation to the outcome of impacts.
Instead the actual distributions should be used, though these still have many uncertainties, particularly
when considering the early evolution of the Solar system when the bombardment was greatest \citep[e.g.,][]{Morbidelli2018}.
Thus the above discussion
should be taken as a caution that the outcome will depend on what is assumed about the relative velocities and size distribution
(and moreover the prescription for the outcome of collisions) and these all contribute to any differences in conclusions
between different authors.
For example, \citet{deNiem2012} concluded that both Earth and Mars atmospheres should grow during the Late Heavy Bombardment,
with 300-600\% growth for the Earth.
Their size distributions are close to a power law with $\alpha=3$ for the cometary population up to $D_{\rm max}=100$\,km,
but are more complex for asteroids (see their fig. 5), while their distribution of impact velocities $\xi$ have means
close to 0.5 and 0.8 for asteroids and comets, respectively (see their figs. 6 and 7).
These still do not explain the different conclusions which must come down to the assumptions about the outcome of
impacts, in particular the optimistic assumptions about impactor retention and the role of giant impacts discussed in
\S \ref{ss:stochasticity}.
Indeed, other authors also find atmospheric loss in impacts \citep{Zahnle1993, Svetsov2007, Pham2011, Pham2016}.

While it remains challenging to make accurate predictions for any given planet, the model can still be used
to make predictions for trends that may be observable in large samples of planets (see \S \ref{ss:pop}).

\subsection{Predictions for exoplanet population}
\label{ss:pop}
\begin{figure*}
  \begin{center}
    \vspace{-0.05in}
    \begin{tabular}{cc}
      \hspace{-0.45in} \includegraphics[width=1.25\columnwidth]{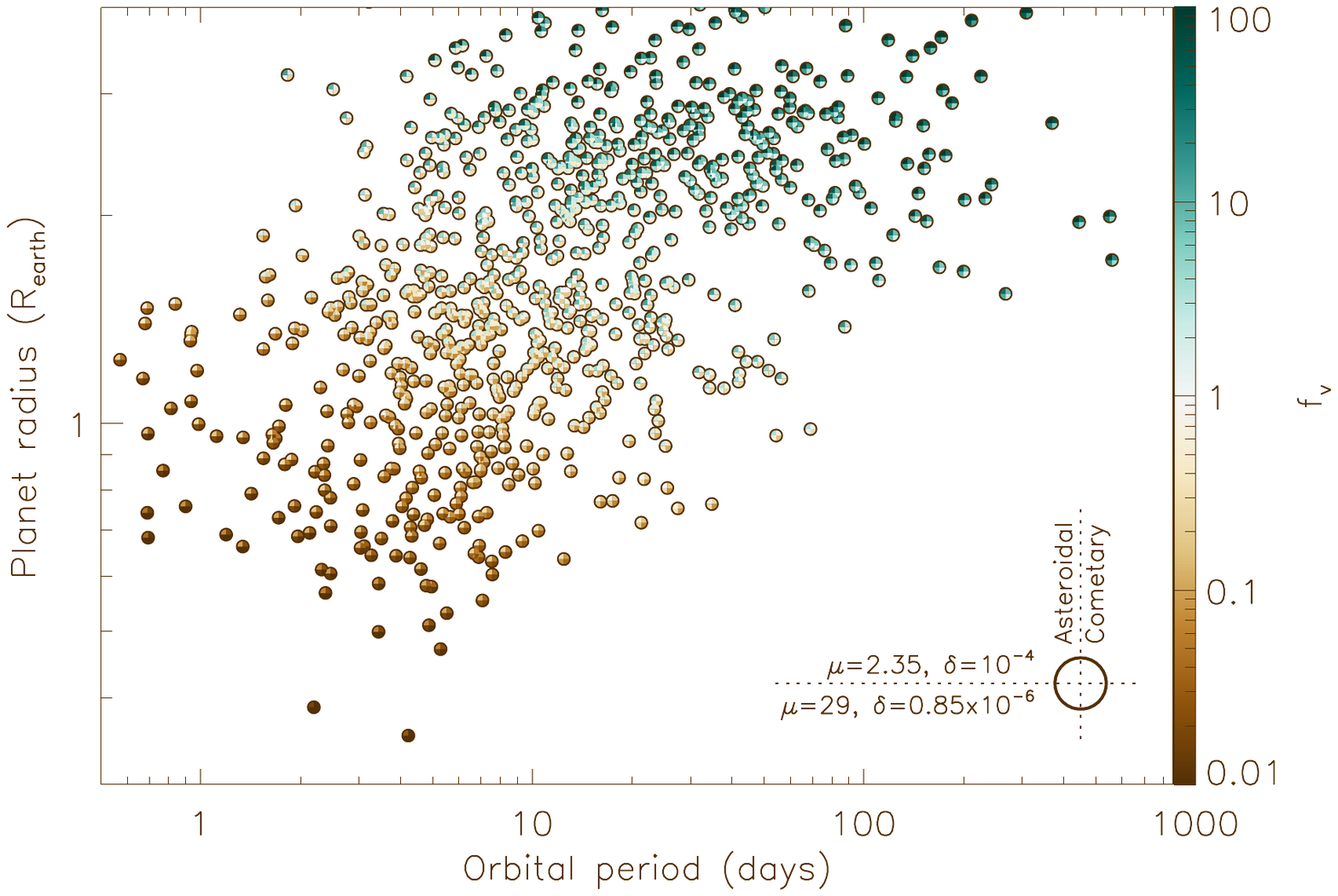} &
      \hspace{-0.85in} \includegraphics[width=1.25\columnwidth]{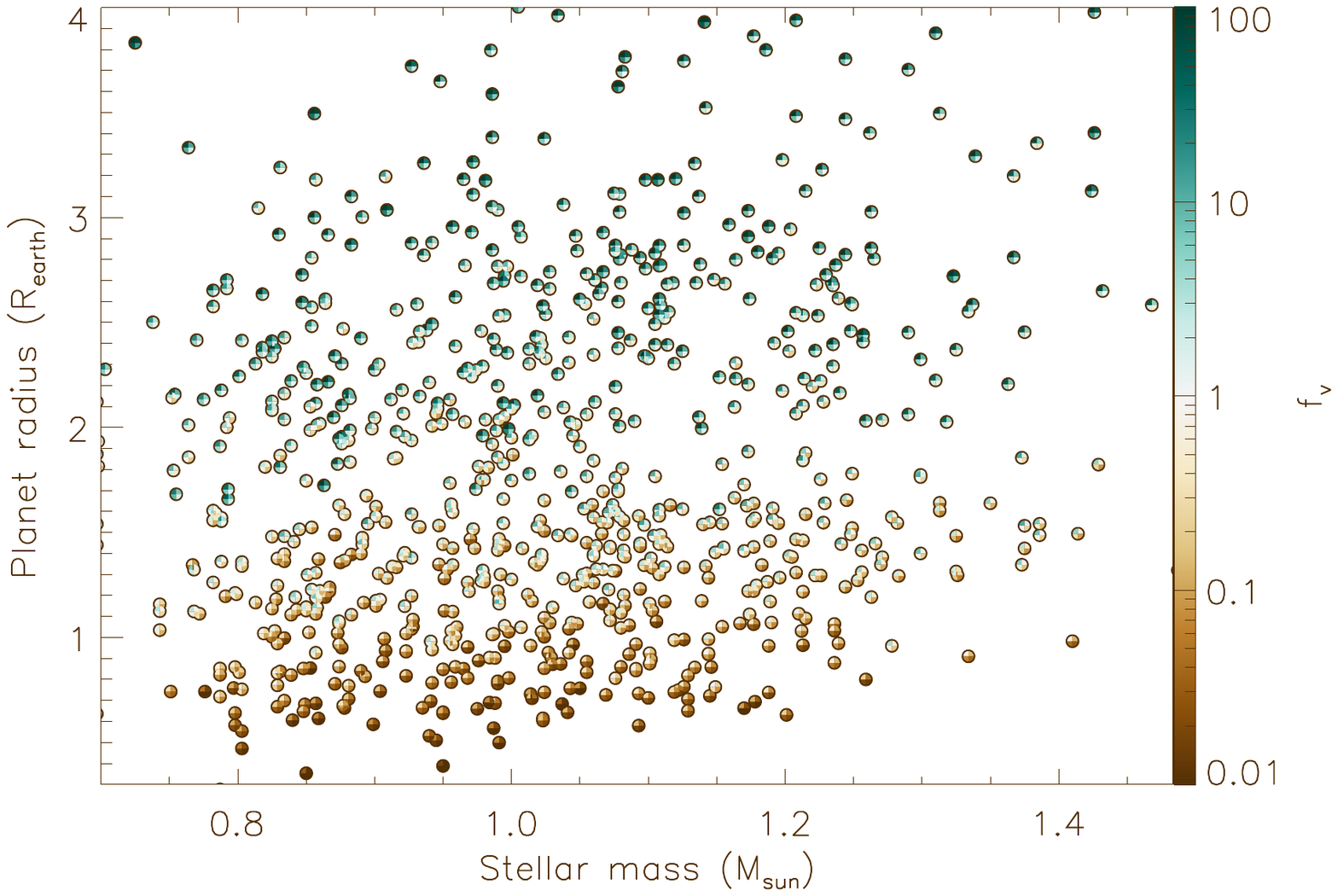} \\[-3.05in]
      \hspace{-0.45in} \includegraphics[width=1.25\columnwidth]{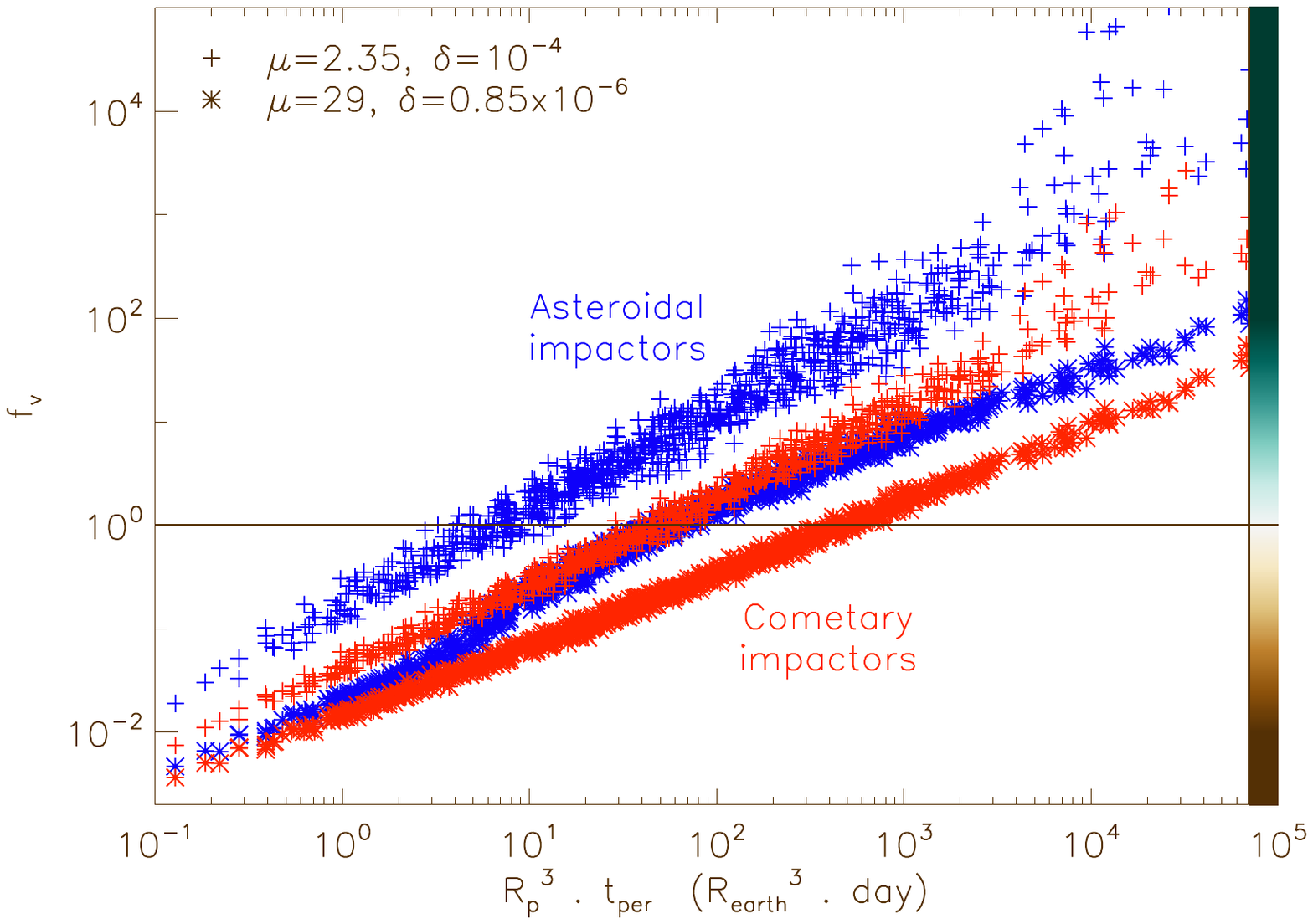} &
      \hspace{-0.85in} \includegraphics[width=1.25\columnwidth]{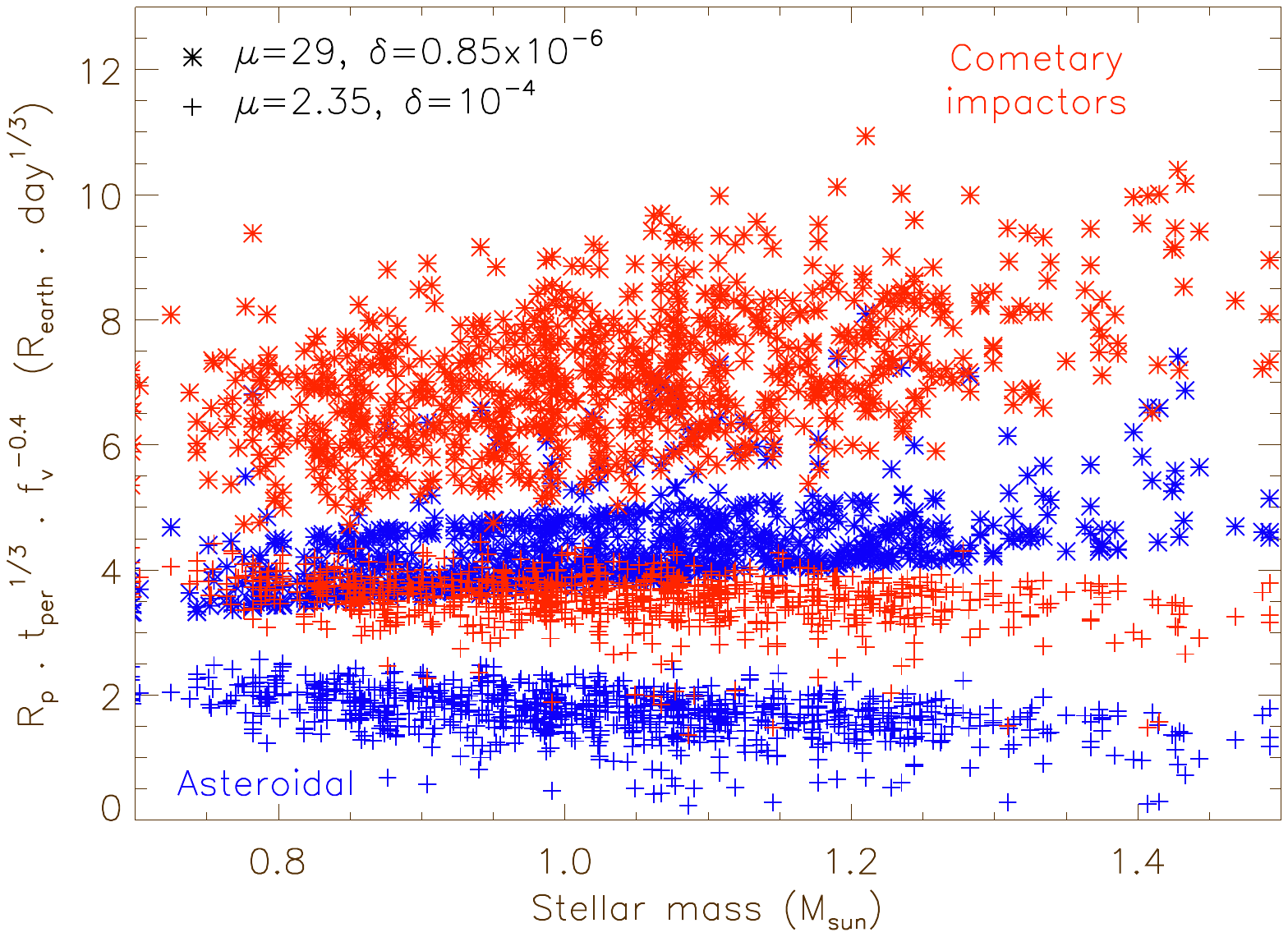}
    \end{tabular}
    \vspace{-2.55in}
    \caption{Model predictions for the population of 907 exoplanets from table 4 of \citet{Fulton2018}.
    The top two plots show planet radius versus either orbital period (top left) or stellar mass (top right),
    and so are respectively equivalent to figs 4 and 8 of \citet{Fulton2018}. 
    For each planet the colour shows the model prediction for $f_{\rm v}$ as indicated in the colour bar on
    the right (i.e., blue is $f_{\rm v}>1$ meaning the atmosphere grows in impacts, brown is $f_{\rm v}<1$ meaning
    the atmosphere depletes in impacts).
    The planets are assumed to have a density 5.5\,g\,cm$^{-3}$, and
    the predictions are shown for four different further assumptions about the impactors or atmosphere, by dividing
    each planet's circle into four quadrants corresponding to the assumptions summarised
    in the bottom right of the top left plot;
    i.e., impactors are assumed to be asteroidal ($\rho_{\rm imp}=2.8$\,g\,cm$^{-3}$, $p_{\rm v}=0.02$) for the left quadrants
    and cometary ($\rho_{\rm imp}=0.9$\,g\,cm$^{-3}$, $p_{\rm v}=0.2$) for the right quadrants,
    the atmosphere is assumed to be Earth-like ($\delta=0.85\times10^{-6}$, $\mu=29$) for the bottom quadrants
    and primordial ($\delta=10^{-4}$, $\mu=2.35$) for the top quadrants.
    The bottom plots show the model predictions for the four different assumptions identified by the colour
    (blue for asteroidal impactors, red for cometary impactors) and symbol (asterisk for Earth-like atmosphere,
    plus for primordial atmosphere).}
   \label{fig:fp18}
  \end{center}
\end{figure*}

Fig.~\ref{fig:fp18} shows the population of exoplanets discovered by Kepler then subsequently followed up
by the California Kepler Survey to determine their accurate radii \citep{Fulton2018}.
In the top left of Fig.~\ref{fig:fp18} the gap in this population, where there is a dearth of transiting exoplanets with radii 
$\sim 1.5R_\oplus$ is evident.
This is interpreted by various authors as evidence of photoevaporation of primordial atmospheres, since it is
only those that are sufficiently large that can survive the bombardment of high energy radiation from the stars shortly
after they reach the main sequence \citep{Owen2017},
although other explanations have been proposed such as the atmospheric mass loss being caused by the
luminosity of the cooling rocky core \citep{Ginzburg2018}.

It is not the purpose of this section to advocate yet another explanation, rather to consider the possible effect
of planetesimal bombardment on the atmospheres in this observed exoplanet population, and so to determine whether
this may have any consequence for their observable properties.
Such consideration faces an obstacle, however, since while the radii and orbital periods of these planets have
been measured with high accuracy, and their stellar properties reasonably well constrained, the masses of the planets
are unknown.
Thus for this analysis it will be assumed that the planets have density of 5.5\,g\,cm$^{-3}$, and so this addresses
the question of how their atmospheres would evolve if they are rocky and their atmospheres contribute little to the
observed radius (which has been the assumption throughout this paper), even though this is not thought to be the case
for the $\gg 1.5R_\oplus$ planets \citep{Rogers2015}.

For each planet, the model is used to predict the $f_{\rm v}$ parameter that determines whether the atmosphere will
grow or deplete in planetesimal impacts for different assumptions about the atmosphere properties (i.e., its mean molecular
weight $\mu$ and fractional mass $\delta$) and about the impactor properties (asteroidal or cometary as defined earlier).
The different quadrants of the circles shown for each planet are for different combinations of these properties.
It is not necessary to focus on the individual quadrants to get the sense that should be clear from the earlier
discussion that the atmospheres of planets towards the top right of the plot are more likely to grow in impacts
(i.e., have a bluer colour and so $f_{\rm v}>1$) while those of planets toward the bottom left of the plot
are more likely to deplete in impacts (i.e., have a redder colour and so $f_{\rm v}<1$).
As discussed previously, the transition between growth and depletion (i.e., the impact shoreline
where planets are coloured in white and so have $f_{\rm v}=1$) depends on the model assumptions.
However, since the most important parameter in the model is the ratio of the planet's escape velocity to its Keplerian
velocity, for each set of assumptions the predicted $f_{\rm v}$ depends mostly on the combination
$R_{\rm p}^3.t_{\rm per}$, where $t_{\rm per}$ is the orbital period, as shown in the bottom left of Fig.~\ref{fig:fp18}.
Fitting a power law for each model shows that $f_{\rm v} \propto [R_{\rm p}^3.t_{\rm per}]^n$, where $n$
is in the range $0.7-1$ for the 4 assumptions shown.

It is noticeable that the planets that are below the gap have atmospheres that are predicted to be depleted in impacts, while
those above the gap are predicted to grow secondary atmospheres in impacts. 
While plotting the observations in this way is not sufficient to extract information about the shape of the gap, for
which consideration of the observational biases is required, such consideration shows that the
radius of the planet at which the gap appears decreases with orbital period \citep{VanEylen2018, Fulton2018}.
The same is true for the transition in the model between atmospheres that grow and deplete, i.e. the impact shoreline which
from the bottom left plot of Fig.~\ref{fig:fp18} is at a radius that scales $R_{\rm p} \propto t_{\rm per}^{-1/3}$.
This consideration also shows that the observed gap is at larger planet radius for planets orbiting higher mass stars, which
can be seen in the top right of Fig.~\ref{fig:fp18}. 
The trend in the model predictions in this regard is less obvious from the top right panel, so this is considered further
in the bottom right panel in which the general trend of the bottom left panel has been removed by assuming $n=0.84$ and
so plotting $R_{\rm p}.t_{\rm per}^{1/3}.f_{\rm v}^{-0.4}$ against stellar mass.
This allows to seek for an additional stellar mass dependence (i.e., in addition to that arising through the
orbital period) of the form
$f_{\rm v} \propto [R_{\rm p}^3.t_{\rm per}]^n M_\star^\gamma$, since the plotted value would be
$\propto M_\star^{-\gamma/(3n)}$ and so flat for $\gamma=0$.
The plotted value can also be used to assess the planet radius at which the $f_{\rm v}=1$ transition would occur
for a fixed orbital period, and shows that for models with Earth-like atmospheres this would appear at larger
planetary radii for higher mass stars (like the trend for the observed gap).
However, the opposite is true for models with more massive primordial atmospheres.

While the model trends show some similarities to the observed properties of the gap it should be cautioned that this does not mean
that planetesimal bombardment would reproduce the observations \citep[e.g.,][]{Lopez2018}.
For example, this application pushes the model into a regime where its assumption that the atmospheres are low in mass
breaks down, and any observable consequence on the properties of the population may require an unrealistic level of
planetesimal bombardment.
The most secure way of interpreting the model predictions in Fig.~\ref{fig:fp18} is to consider the effect of bombardment on
a planet that is born with a low mass (e.g., Earth-like) atmosphere.
The prediction is that planets below the gap would find it hard to grow a secondary atmosphere due to impacts.
However, since more massive atmospheres have a larger $f_{\rm v}$, if they do start to grow an atmosphere then this likely
becomes easier, but this does not address the question of whether the planet can grow an atmosphere that is massive
enough to become inflated and so change its position on the plot and so be responsible for the gap. 
That would depend on the amount accreted and on how the physics changes as the atmosphere becomes more massive,
for example the higher mean molecular weight of a secondary atmosphere could mean that a significantly higher
fraction of the planet's mass than a few \% is required to be accreted for it to appear inflated
(e.g., by a factor of $\sim \mu_\oplus/\mu_\odot \approx 12$).
But if the current model were applicable to more massive atmospheres, its predictions for atmosphere growth of
$\Delta m/m_{\rm ac}$ of a few \% (see Fig.~\ref{fig:outcome1}) would suggest that bombardment levels comparable with the planet
mass are required to attain an atmosphere of a few \%.

The prediction that planets below the gap cannot grow secondary atmospheres by impacts also applies to planets
that may have lost their atmosphere due to photoevaporation, since that may be the origin of the gap and bombardment
may continue after that process is complete.
Thus it is worth noting that the prediction is to some extent dependent on the assumptions about the impacts, so that
planets just below the gap may be able to grow secondary atmospheres if the impact conditions are right
(i.e., some of the planets below the gap have quadrants that are light blue in Fig.~\ref{fig:fp18}).
Depending on the exact slope of the gap, it could be that planets at larger distance from the star are more amenable to
growth of impact-generated secondary atmospheres (following loss of their primordial atmospheres by photoevaporation).

The interpretation of the predictions for the effect of bombardment on a planet that is born with a massive atmosphere are less
secure.
However, this shows that for planets below the gap such atmospheres would be expected to be depleted, though of course only
if sufficient bombardment occurs.
As above, if the current model were applicable to more massive atmospheres, its prediction for atmosphere
loss of $\Delta m/m_{\rm ac}$ of order 1\% (see Fig.~\ref{fig:outcome1}) would suggest that bombardment levels
comparable with the planet mass would be required to remove a few \% atmosphere.
For planets above the gap, their atmospheres would be expected to grow in impacts, and to become more volatile-rich.
If future observations show their atmospheres to be volatile-rich then this model would support planetesimal impacts
being one possible origin for the volatiles.
It must, however, be noted that volatile-rich atmospheres may also be replenished by outgassing
\citep[as may be the case for Mars for example,][]{Craddock2009}, a process that is
not considered here.

\subsection{Implications for life}
\label{ss:life}
With the origin of life on Earth still debated, uncertainty
in extrapolating to other planetary systems is unavoidable.
However, impacts are often considered to play a positive role, for example by delivery of organic molecules or their synthesis
in impact shocks \citep{Chyba1992, Patel2015}, or by the delivery of water to otherwise dry planets \citep[e.g.,][]{Chyba1990a}.
Though impacts may also inhibit the further development of life \citep{Maher1988}.
Since the Earth's evolution was evidently conducive to the development of life,
then if we make the anthropocentric assumption that a similar evolution in terms of a planet's atmosphere might be similarly
conducive to life, the results from this paper can be used to make relative statements about whether planets in the habitable
zones of other stars would be more or less conducive to the development of life.

\begin{figure}
  \begin{center}
    \vspace{-0.6in}
    \begin{tabular}{c}
      \hspace{-0.55in} \includegraphics[width=1.25\columnwidth]{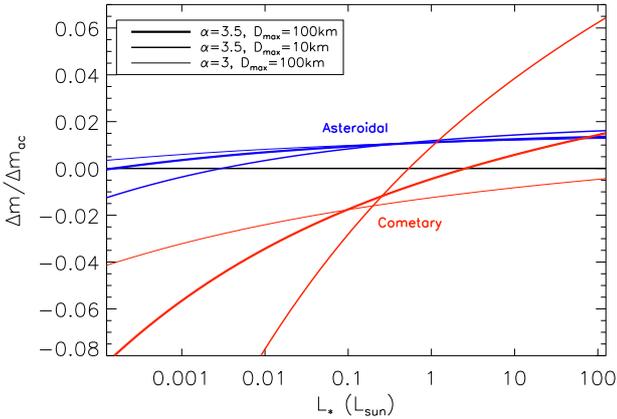}
    \end{tabular}
    \vspace{-2.55in}
    \caption{Change in atmosphere mass per accreted impactor mass for Earth-like planets ($1{\rm M}_\oplus$ with a $0.85 \times 10^{-6}{\rm M}_\oplus$
    atmosphere with $\mu=29$) in the habitable zone
    of stars of different luminosity (i.e., $a_{\rm p}=\sqrt{L_\star}$, assuming $M_\star \propto L_\star^{1/3}$). 
    This is plotted for different assumptions about the impactors.
    The size distribution is assumed to be a power law from $D_{\rm min}=1$\,m up to $D_{\rm max}=10$\,km or 100\,km,
    with a slope of $\alpha=3.0$ or 3.5.
    Asteroidal impactors are those with $\rho_{\rm imp}=2.8$\,g\,cm$^{-3}$, $p_{\rm v}=0.02$ and $\xi=0.3$,
    and cometary impactors are those with $\rho_{\rm imp}=0.9$\,g\,cm$^{-3}$, $p_{\rm v}=0.2$ and $\xi=1.0$.
    }
   \label{fig:dmmdmhz}
  \end{center}
\end{figure}

Fig.~\ref{fig:dmmdmhz} shows the change in atmosphere mass for an Earth-like planet in the habitable zone
of stars of different luminosity.
Here it has been assumed that $L_\star=M_\star^3$ (for units of $L_\odot$ and $M_\odot$), and the
habitable zone is simply taken as the distance at which its temperature is 278\,K so that $a_{\rm p}=\sqrt{L_\star}$
\citep[see e.g.][for a more detailed definition]{Kopparapu2014}.
It then considers the fractional change in the atmosphere for different assumptions about the impacting planetesimals.
This shows that there is a general tendency for habitable planets around lower luminosity stars to be more susceptible to having their
atmospheres depleted in collisions,
which is true regardless of the assumption about the impacting planetesimals.
This is because the habitable zone is closer in for lower luminosity stars, which even when accounting for the slower orbital
velocity due to the lower stellar mass, results in higher collision velocities and
so more destructive impacts (for the given assumptions the collision velocity in the habitable zone scales
$\propto M_\star^{-1/4}$).

There is already much discussion about the habitability of planets around low mass M stars \citep[e.g.,][]{Shields2016}, since 
close-in planetary systems are common around such stars, 
and the proximity of the habitable zone to low luminosity stars makes these planets relatively
easy to detect and further characterise using transit observations \citep[e.g.,][]{deWit2018}.
However, it was shown that such planets that end up in the habitable zone would have exceeded the runaway greenhouse threshold
on the pre-main sequence and so would have lost any water \citep{Ramirez2014}, which is confounded by issues such
as the high incidence of flares on low mass stars that would be detrimental to habitability \citep{Vida2017, Tilley2019},
and the likelihood of these habitable zone planets to be tidally locked to the host star with consequences for
atmospheric dynamics \citep{Kopparapu2016}. 
Impacts could provide a potential solution to some of these issues, by delivering a secondary atmosphere and water to
the planets.
However, Fig.~\ref{fig:dmmdmhz} shows that, at least as long as the impacting planetesimals have similar properties to
those hitting the Earth, impacts are more likely to destroy the atmosphere of a habitable zone planet around a low mass star
than to replenish it.
Fortunately the impacting planetesimals may have a different impact velocity distribution, so that habitable zone
planets could still grow substantial atmospheres as was found for the TRAPPIST-1 planets by \citet{Kral2018}.

In any case, it might be noted that planets in the habitable zones of higher mass stars may be more susceptible
to the growth of a secondary atmosphere in impacts.
Although the fact that the lines are relatively flat on Fig.~\ref{fig:dmmdmhz} (at least for certain assumptions)
could also be taken to infer that the atmospheres of Earth-like
habitable zone planets do not suffer significantly different fates to the Earth as a result of impacts.
However, a strong conclusion on this would require knowledge of the possible impacting planetesimal population, which
may be systematically different around stars of different spectral type.
There is also the caveat that water could be retained in the magma ocean during formation and outgassed
later on \citep{Peslier2017, Ikoma2018}, so that an Earth-like impact history may not be a necessary requirement for
the development of life.

\section{Conclusion}
\label{s:conc}
This paper has developed a model for the evolution of planetary atmospheres due to planetesimal impacts that
accounts for both stripping of the atmosphere and the delivery of volatiles.
It is based on a suite of simulations of impacts that covers a wide range of planetary atmosphere and
impacting planetesimal properties.
The implications of the model for the atmosphere evolution of planets in different regions of parameter space
is discussed, and the relative simplicity of the parameterisation means that it is possible to understand both
qualitatively and quantitatively the dependence of the outcome on the different input parameters (i.e.,
the impacting planetesimals' densities, volatile fractions and impact velocities, as well as the planet
mass, orbital distance and atmospheric mass and composition, and the stellar properties).

The conclusion is that planets are divided in planet mass vs semimajor axis parameter space into those with
atmospheres that deplete in impacts (if they are close to the star and/or low in mass) and those that can
grow secondary volatile-rich atmospheres (if they are far from the star and/or high in mass).
The dividing line, or impact shoreline, is parallel to one of constant ratio of orbital velocity to escape velocity, and is analogous to
the cosmic shoreline discussed in \citet{Zahnle2017} that was interpreted as a consequence of irradiation.
The location of the impact shoreline depends on assumptions about impacting planetesimals,
and for different (reasonable) assumptions there is more than an order of magnitude spread,
say in terms of its location in planet mass for a given orbital distance.
For Sun-like stars, a planet with properties like the Earth would sit near the shoreline.

Impact driven atmosphere evolution is dominated by the combined effect of accreting 1-20\,km planetesimals,
so as long as the size distribution extends beyond this range, the conclusions are largely independent of the
size distribution.
However, the model presented herein is based on simulations appropriate for low mass atmospheres, and further
development is needed to consider the situation for massive atmospheres for which such planetesimals would
undergo an aerial burst (rather than be destroyed on reaching the planet surface).
As in previous studies, giant impacts are found to have little effect on atmosphere evolution unless the atmosphere is a
significant fraction of the planet mass, though they may introduce an element of stochasticity when impactors are
comparable in mass to the planet.

Applying the model to the Solar system terrestrial planets shows that whether the Earth's atmosphere grows
or depletes in impacts is strongly dependent on the distribution of impact velocities and impactor properties.
Further discussion of this is deferred to a later paper where these distributions can be considered in more detail.

Application to the population of transiting exoplanets discovered by Kepler shows that the gap in the planet radius
distribution is roughly coincident with the dividing line (impact shoreline) between planets with atmospheres
that grow and deplete in collisions.
The dependence of this dividing line on orbital distance and stellar mass is also similar to that observed.
It seems unlikely that bombardment levels would have been sufficient to be responsible for the gap,
either by depleting the primordial atmospheres of the smallest planets, 
or by growing substantial secondary atmospheres for the most massive planets,
since this would require bombardment by a mass comparable to the planets
(and even such high bombardment levels may not be sufficient).
However, it must be remembered that the predictions of the model are inaccurate for
planets with atmospheres as massive as those inferred for planets above the gap (i.e., a few \% of the planet mass).
Nevertheless, this coincidence shows that the effect of impacts onto planetary atmospheres deserves further consideration.
It is also possible to draw firmer conclusions about planets below the gap, for example, that if these atmospheres
were depleted by stellar irradiation, then they would be unlikely to grow a secondary atmosphere in impacts,
except for those just below the gap and for certain conditions on the impacting planetesimals.
Consideration of planets in the habitable zone of stars of different mass shows that impacts are more harmful
for those of lower mass stars \citep[see also][]{Kral2018}. 
Thus if an Earth-like bombardment, and its effect on the Earth's atmosphere, was a requirement for the development
of life, this may give cause to disfavour M stars as the hosts of life-bearing planets.
However, without consideration of the impactor populations, or of the other factors relevant to the evolution
of the conditions on the planetary surface, this cannot be a strong conclusion.

\appendix
\section{Parameter summary}

\begin{table*}
  \centering
  \caption{Summary of parameters used in the paper and their units.}
  \label{tab:units}
  \begin{tabular}{lll}
     \hline
     Parameter                & Units                 & Meaning \\
     \hline 
     $a_{\rm p}$              & au                    & Planet semimajor axis \\
     $D$                      & m                     & Impactor diameter \\
     $D_{\rm GI}$             & m                     & Impactor diameter above which giant impacts dominate atmosphere mass loss \\
     $D_{\rm min}$            & m                     & Minimum impactor diameter \\
     $D_{\rm max}$            & m                     & Maximum impactor diameter \\
     $f_{\rm v}$              &                       & Ratio of gain of atmosphere mass due to impactor retention to mass loss in impacts \\
     $H$                      & m                     & Atmospheric scale height \\
     $L_\star$                & ${\rm L}_\odot$       & Stellar luminosity \\
     $M_\star$                & ${\rm M}_\odot$       & Stellar mass \\
     $M_{\rm p}$              & ${\rm M}_\oplus$      & Planet mass \\
     $m$                      & ${\rm M}_\oplus$      & Total atmosphere mass \\
     $\dot{m}^{-}$            & ${\rm M}_\oplus$\,s$^{-1}$  & Atmospheric mass loss rate \\
     $\dot{m}_{\rm v}^{+}$    & ${\rm M}_\oplus$\,s$^{-1}$  & Rate at which atmosphere gains volatiles due to impactor retention \\ 
     $m_{\rm inc}$            & ${\rm M}_\oplus$      & Total mass of impactors put on planet crossing orbits \\
     $m_{\rm ac}$             & ${\rm M}_\oplus$      & Total mass of impactors accreted by the planet \\
     $m_{\rm atmloss}(D)$     & ${\rm M}_\oplus$      & Atmospheric mass lost in impact with impactor of diameter $D$ \\
     $m_{\rm impacc}(D)$      & ${\rm M}_\oplus$      & Mass of impactor of diameter $D$ that is retained by planet \\
     $m_{\rm atmloss}$        & ${\rm M}_\oplus$      & Atmospheric mass lost integrated over the impactor size distribution \\
     $m_{\rm atmlss,GI}$      & ${\rm M}_\oplus$      & Atmospheric mass lost by giant impacts integrated over the impactor size distribution \\
     $m_{\rm impacc}$         & ${\rm M}_\oplus$      & Impactor mass retained by planet integrated over the impactor size distribution \\
     $m_0$                    & ${\rm M}_\oplus$      & Total initial atmosphere mass \\
     $m_{\rm p}$              & ${\rm M}_\oplus$      & Mass of primordial component of atmosphere \\
     $m_{\rm v}$              & ${\rm M}_\oplus$      & Mass of volatile (secondary) component of atmosphere \\
     $m_{\rm imp}$            & ${\rm M}_\oplus$      & Mass of impacting planetesimal \\
     $n(D)dD$                 &                       & Number of impactors in size range $D$ to $D+dD$ \\
     $p_{\rm v}$              &                       & Fraction of retained impactor mass that goes into the atmosphere \\
     $q$                      & au                    & Pericentre distance of impactor orbit \\
     $R_{\rm ac}$             & s$^{-1}$              & Rate at which impactors collide with the planet \\
     $R_{\rm ej}$             & s$^{-1}$              & Rate at which impactors are ejected by the planet \\
     $R_{\rm dyn}$            & s$^{-1}$              & Rate at which impactors are removed dynamically from planet-crossing orbits \\
     $R_{\rm p}$              & m                     & Planet radius \\
     $T$                      & K                     & Temperature of planet atmosphere \\
     $t$                      & s                     & Time \\
     $t_0$                    & s                     & Time for atmosphere to deplete in absence of volatile replenishment, $t_0=m_0/\dot{m}_0^{-}$ \\
     $t_{\rm bare}$           & s                     & Time for atmosphere to be completely depleted \\
     $t_{\rm per}$            & day                   & Orbital period \\
     $v_{\rm imp}$            & m\,s$^{-1}$           & Impact velocity \\
     $v_{\rm p}$              & m\,s$^{-1}$           & Planet's orbital velocity \\
     $v_{\rm esc}$            & m\,s$^{-1}$           & Planet's escape velocity \\
     $x$                      &                       & Parameter equal to $(m_{\rm imp}/M_{\rm p})(v_{\rm imp}/v_{\rm esc})$ \\
     $\alpha$                 &                       & Power law index of impactor size distribution \\
     $\delta$                 &                       & Ratio of atmosphere mass to planet mass \\
     $\delta_0$               &                       & Ratio of initial atmosphere mass to planet mass \\
     $\delta_{\rm GI}$        &                       & Ratio of atmosphere to planet mass above which giant impacts dominate evolution \\
     $\eta$                   &                       & Parameter that for a given planet and scenario scales with impactor size cubed \\
     $\eta_{\rm ab}$          &                       & Defines the smallest planetesimal that does not undergo aerial burst before impact \\
     $\eta_{\rm maxret}$      &                       & Defines the largest impacting planetesimal whose mass can be retained by the planet \\
     $\eta_{\rm tr}$          &                       & Defines the smallest planetesimal that does not fragment in atmosphere before impact \\
     $\mu$                    &                       & Mean molecular weight of atmosphere \\
     $\xi$                    &                       & Averaged ratio of planet-impactor relative velocity to planet orbital velocity \\
     $\rho_0$                 & g\,cm$^{-3}$          & Atmosphere density at surface \\
     $\rho_{\rm p}$           & g\,cm$^{-3}$          & Planet density \\
     $\rho_{\rm ps}$          & g\,cm$^{-3}$          & Density of the planetary surface \\
     $\rho_{\rm imp}$         & g\,cm$^{-3}$          & Impactor density \\
     $\chi_{\rm a}$           &                       & Parameter used to determine atmospheric mass loss in collision \\
     $\chi_{\rm pr}$          &                       & Parameter used to determine impactor retention in collision \\
     \hline
  \end{tabular}
\end{table*}

\bibliographystyle{mnras}
\bibliography{refs.bib} 

\bsp    
\label{lastpage}
\end{document}